\documentclass[]{aa}  

\DeclareUnicodeCharacter{FEFF}{}
\usepackage[]{natbib}
\usepackage[T1]{fontenc}
\usepackage{ae,aecompl}
\usepackage[dvipsnames]{xcolor}
\usepackage{psfrag,color}
\usepackage[]{inputenc}
\usepackage{graphicx}	
\usepackage{amsmath}	
\usepackage{amssymb}	
\usepackage{longtable}
\usepackage{adjustbox}
\usepackage{ulem}       
\usepackage{newtxmath} 
\usepackage{newtxtext}
\usepackage{pdflscape}
\usepackage{supertabular}
\usepackage{rotating}
\usepackage{footnote}
\usepackage{times}
\usepackage{chemmacros}
\usepackage{xspace}
\usepackage{subcaption}

\usepackage[colorlinks=true, urlcolor=blue, linkcolor=blue, citecolor=blue]{hyperref} 

\definecolor{lime}{HTML}{A6CE39}
\DeclareRobustCommand{\orcidicon}{
	\begin{tikzpicture}
	\draw[lime, fill=lime] (0,0) 
	circle [radius=0.16] 
	node[white] {{\fontfamily{qag}\selectfont \tiny ID}};
	\draw[white, fill=white] (-0.0625,0.095) 
	circle [radius=0.007];
	\end{tikzpicture}
	\hspace{-2mm}
}

\foreach \x in {A, ..., Z}{\expandafter\xdef\csname orcid\x\endcsname{\noexpand\href{https://orcid.org/\csname orcidauthor\x\endcsname}
			{\noexpand\orcidicon}}
}

\usepackage[]{ulem}  
\newcommand\redout{\bgroup\markoverwith
{\textcolor{red}{\rule[0.5ex]{2pt}{0.8pt}}}\ULon}

\newcommand{\hii}{H\,\textsc{ii}}
\newcommand{\heii}{He\,\textsc{ii}}
\newcommand{\cii}{[C\,\textsc{ii}]}
\newcommand{\oii}{[O\,\textsc{ii}]}
\newcommand{\oiii}{[O\,\textsc{iii}]}
\newcommand{\oiv}{[O\,\textsc{iv}]}

\newcommand{\nii}{[N\,\textsc{ii}]}

\newcommand{\nev}{[Ne\,\textsc{v}]}

\newcommand{\sii}{[S\,\textsc{ii}]}
\newcommand{\siii}{[S\,\textsc{iii}]}

\newcommand{\ariv}{[Ar\,\textsc{iv}]}
\newcommand{\arv}{[Ar\,\textsc{v}]}

\newcommand{\feiii}{[Fe\,\textsc{iii}]}

\newcommand{\nel}{$n_{\rm e}$} 
\newcommand{\tel}{$T_{\rm e}$}

\begin{document}

\title{Investigating the electron temperature of {\ariv} in planetary nebulae using the DESIRED database.}

   \subtitle{}

   \author{J. Garc\'{\i}a-Rojas\inst{1,2}{\orcidB{}} 
          \and
          E. Reyes-Rodr\'iguez\inst{1,2}{\orcidA{}}
          \and
          J. E. M\'endez-Delgado\inst{3}{\orcidC{}}
          \and
          C. Morisset\inst{4,5}{\orcidD{}}
          \and
          D. Jones\inst{1,2}{\orcidI{}}
          \and
          C. Esteban\inst{1,2}{\orcidD{}}
          \and
          F. F. Rosales-Ortega\inst{6}{\orcidJ{}}
          \and
          V. G\'omez-Llanos\inst{1,2}{\orcidF{}}
          \and
          M. Orte-Garc\'{\i}a\inst{2,3}{\orcidG{}}
          \and
          L. E. Mart\'inez-Rivero\inst{3}{\orcidM{}}
          \and
          Y. Hong\inst{7,8}{\orcidK{}}
          \and
          X. Fang\inst{7,8,9,10}{\orcidL{}}
     }

   \institute{Instituto de Astrof\'isica de Canarias, E-38205 La Laguna, Tenerife, Spain 
   \email{jogarcia@iac.es}
        \and
        Departamento de Astrof\'isica, Universidad de La Laguna, E-38206 La Laguna, Tenerife, Spain   
        \and             
        Instituto de Astronom\'{\i}a, Universidad Nacional Aut\'onoma de M\'exico, Ap. 70-264, 04510 CDMX, Mexico
        \and
        Instituto de Astronom\'{\i}a, Universidad Nacional Aut\'onoma de M\'exico, Apdo. Postal 877, 22800, Ensenada, B.C., Mexico
        \and
        Instituto de Ciencias F\'{\i}sicas, Universidad Nacional Aut\'onoma de M\'exico, Av. Universidad s/n, 62210 Cuernavaca, Morelos, Mexico
        \and
        Instituto Nacional de Astrofísica, Óptica y Electrónica (INAOE SECIHTI), Luis E. Erro 1, 72840, Tonantzintla, Puebla, Mexico
        \and
        CAS Key Laboratory of Optical Astronomy, National Astronomical Observatories, Chinese Academy of Sciences (NAOC), Beijing 100101, People's Republic of China
        \and
        School of Astronomy and Space Sciences, University of Chinese Academy of Science (UCAS), Beijing 100049, People's Republic of China
        \and
        Xinjiang Astronomical Observatory, Chinese Academy of Sciences, 150 Science 1-Street, Urumqi, Xinjiang 830011, People's Republic of China
        \and
        Laboratory for Space Research, Faculty of Science, The University of Hong Kong, Pokfulam Road, Hong Kong, People's Republic of China
        }

   \date{Received ; accepted }

\authorrunning {Garc\'ia-Rojas et al.}
\titlerunning {$T_{\rm e}$([Ar~{\sc iv}]) in planetary nebulae using the DESIRED database}
   \date{\today}

 
  \abstract
   {The determination of the chemical composition of the interstellar medium relies heavily on the analysis of emission lines from ionised nebulae. While most studies focus on low- to mid-ionisation species, the physical conditions in the innermost, high-ionisation regions of planetary nebulae (PNe) remain poorly constrained, particularly regarding the role of additional heating mechanisms.}
   {We investigate the behaviour of the electron temperature derived from {\ariv} lines, {\tel}({\ariv}), in a sample of PNe to characterize the thermal structure of high-excitation gas and compare it with the predictions of standard photoionisation models.}
   {Using the DEep Spectra of ionised REgions Data Base Extended (DESIRED-E), we selected a sample of 57 PNe for which {\tel}({\ariv}), {\tel}({\oiii}), and {\nel}({\ariv}) could be determined simultaneously and homogeneously. We performed a detailed comparison between these observational diagnostics and a suite of over 160,000 photoionisation models from the Mexican Million Models Database (3MdB).}
   {We find that the observed {\tel}({\ariv}) values are systematically higher than those predicted by pure photoionisation models for a given {\tel}({\oiii}), i.e. approximately 31\% of the PNe sample exhibits a \tel({\ariv}) more than $2\sigma$ higher than photoionization model predictions. This discrepancy persists regardless of the specific set of auroral lines used for the diagnostic or the choice of atomic data (transition probabilities and collision strengths) adopted in the calculations. Other high-ionisation \tel\ diagnostics, compiled from the DESIRED datababase or from the literature, however, do not show such behaviour, though the statistics for these are much more limited.}
   {The fact that models succeed for the highest-ionisation species but fail specifically for {\ariv} suggests that the discrepancy is not due to a global inner-nebula heating mechanism. Instead, it points toward a localized physical effect or a limitation in the current understanding of the ionisation stratification and atomic physics specific to the Argon ion stages. This ``{\ariv} anomaly'' must be resolved to ensure the reliability of abundance determinations in high-excitation nebulae.}


\maketitle
%

\section{Introduction}
\label{sec:intro}

The analysis of emission-line spectra from ionised nebulae enables the determination of the chemical composition of the interstellar medium (ISM), both in our Galaxy and in external galaxies. This constitutes a fundamental step towards understanding stellar nucleosynthesis processes, as well as the chemical evolution of galaxies and the Universe as a whole. 

Photoionised nebulae, such as {\hii} regions and planetary nebulae (PNe), are luminous astrophysical objects whose chemical abundances provide crucial insights into ISM enrichment and its evolution over cosmic time. Their optical spectra are dominated by emission lines arising from collisional excitation (CELs) and recombination (RLs). Despite the overwhelming abundance of hydrogen and helium, CELs from heavier elements such as oxygen, nitrogen, neon, and sulphur are prominent, as their ionic structures favour efficient collisional excitation at typical electron temperatures of ionised gas. {\hii} regions and PNe, however, exhibit notable differences: {\hii} regions are large, massive structures ionised by young OB-type stars with effective temperatures in the range of ($T_{\rm eff}$$\sim$25–50 kK), whereas PNe are more compact and less massive, but their central stars can reach much higher temperatures (up to 250 kK).  These extreme temperatures allow the presence of highly ionised species such as {\heii}, {\nev}, and {\arv} \citep[see e.g.,][]{GarciaRojas:15}. Consequently, their spectra and ionisation structures are markedly different, which is also reflected in narrow-band imaging.

A detailed characterisation of the electron temperature ({\tel}) and electron density ({\nel}) distributions within an ionised nebula is essential for the accurate determination of emission-line emissivities and, consequently, elemental abundances. This is particularly critical for CELs, whose emissivities exhibit an exponential dependence on {\tel}, especially in the optical regime. Reliable temperature diagnostics are therefore required to avoid systematic biases in ionic abundance determinations. In nebulae with complex thermal structures and potential temperature inhomogeneities, CEL-based diagnostics tend to overestimate the average {\tel}, as higher-temperature regions disproportionately contribute to the emission of CELs due to the exponential term \citep{Peimbert:67}. This bias can be mitigated by using weak RLs, which are much less sensitive to temperature fluctuations. A consequence of this effect is the so-called ``abundance discrepancy problem'', whereby RL-based ionic abundances are systematically higher than those derived from CELs \citep[see][and references therein]{Peimbert:17, GarciaRojas:20}.

A homogeneous analysis of a statistically significant sample of high-quality spectra is crucial to investigate the impact of {\nel} and {\tel} structures on chemical abundance determinations and metallicity estimates in ionised nebulae. In this context, \citet{MendezDelgado:23a} proposed a plausible solution to the long-standing abundance discrepancy problem in {\hii} regions, attributing it to temperature fluctuations in the high-ionisation zones, potentially driven by stellar feedback. This interpretation suggests that most metallicity estimates based on CELs could be severely underestimated and must be revised accordingly. The effect is expected to be even more pronounced in low-metallicity environments, such as the high-redshift galaxies observed by the James Webb Space Telescope (JWST). However, this explanation does not appear to account for the abundance discrepancies observed in PNe, highlighting the need for further investigation using a well-characterised and homogeneously analysed dataset of PNe.  

PNe are known to exhibit highly heterogeneous ionisation structures owing to the broad spectral energy distributions of their ionising sources. This leads to ionisation structures that differ significantly from those of {\hii} regions, particularly in highly excited PNe. Therefore, a detailed examination of the physical conditions in the innermost regions of PNe --where emission from high-ionisation species originates-- is essential, both observationally and theoretically, to explore the possible contribution of additional excitation mechanisms beyond pure photoionisation.

In this work we investigate the behaviour of the poorly studied electron temperature associated with Ar$^{3+}$, {\tel}({\ariv}), in a large sample of PNe. Despite the similarity between the ionisation potentials (IP) of Ar$^{3+}$ (40.74~eV) and O$^{2+}$ (35.12~eV), their spatial distributions within photoionised nebulae are distinct due to the higher energy required to reach their subsequent ionisation stages. A$r^{3+}$ acts as a higher-excitation species because its existence is tied to a ``harder'' radiation field; specifically, its destruction into Ar$^{4+}$ requires 59.81~eV, placing its peak abundance within the He$^{2+}$ zone (requiring photons with energies $>54.4$ eV to double-ionise He) found in the innermost nebular regions. In contrast, the O$^{2+}$ population is maintained across a broader, lower-energy window and is effectively bounded by the He~{\sc i}/{\sc ii} transition. This is observationally supported by recent PNe Integral Field Unit (IFU) observations, where Ar$^{3+}$ appears to originate in more central regions than O$^{2+}$ \citep{GarciaRojas:22, GomezLlanos:24}, making it a valuable diagnostic of the physical conditions in the inner, high-excitation zones of PNe, closer to the central ionising source.

Historically, the detection of {\ariv} auroral lines has been a challenging task due to their inherent weakness. Early work by \citet{Czyzak:80} noted that the observed {\ariv} auroral lines in a sample of six PNe were systematically too strong --i.e. will provide higher \tel({\ariv})-- relative to the nebular transitions when compared against theoretical predictions of the time. They concluded that the available collision strengths and $A$-values were likely inaccurate. Later, \citet{Keenan:97} attempted to validate the new collision strengths computed by \citet{Ramsbottom:97} using deep, high-resolution spectra of nine PNe. They derived \tel(\ariv) by combining nebular lines with each of the three detected auroral lines (see Sect.~\ref{sec:desired}) and compared these with other diagnostics of similar ionization potential, such as {\oiii} and {\siii}. This comparison assumed that these ions share similar spatial distributions with \ariv, an assumption that is not necessarily valid, as discussed above. While significant discrepancies were found for two objects, \citet{Keenan:97} concluded that this atomic data set yielded \tel(\ariv) values that were mutually consistent across the different auroral lines and in good agreement with other temperature diagnostics. In this work, we revisit these results by significantly increasing the sample of PNe with determined \ariv\ temperatures.

This paper is organised as follows. In Sect.~\ref{sec:desired} we describe the methodology used to derive the physical conditions of the DESIRED PNe sample. Sect.~\ref{sec:results} presents the results obtained from the analysis of the physical conditions diagnostics. In Sect.~\ref{sec:tests} we assess the possible observational and methodological biases that may affect the observed behaviour of the \tel({\ariv}). Finally, Sect.~\ref{sec:discuss} discusses the results, and Sect.~\ref{sec:conclu} summarises our main conclusions.

\section{{\ariv} diagnostic lines in the DESIRED database}
\label{sec:desired}

To address key challenges in nebular astrophysics using a homogeneous and systematic approach, \citet{MendezDelgado:23b} developed the DEep Spectra of Ionised REgions Data Base (DESIRED) project, which has been extended since then in the DESIRED extended \citep[DESIRED-E;][]{MendezDelgado:24} database. This resource compiles hundreds of intermediate- to high-spectral-resolution optical spectra of photoionised nebulae. The main strategy for building DESIRED-E has prioritised the measurement of auroral CELs sensitive to electron temperature, alongside multiple CEL-based density diagnostics. A significant fraction of the compiled data were obtained with 8–10 m class telescopes and optimised to detect faint emission features. To compute the physical conditions and chemical abundances in the DESIRED-E photoionised nebulae\footnote{As of March 2026, the DESIRED-E database includes spectra of 386 Galactic and extragalactic planetary nebulae, and a total number of 3154 photoionised nebulae spectra.}, we use {\sc PyNeb} v.~1.1.30 \citep{Luridiana:15, Morisset:20, Mendoza:23}.

\begin{table}[h!]
\begin{center}
\caption{PNe for which \tel({\ariv}), \tel({\oiii}) and \nel({\ariv}) were determined simultaneously, fulfilling the criteria described in Sect.~\ref{sec:desired}. The subsamples indicate the temperature-sensitive lines used to compute {\tel}({\ariv}).}
\label{tab:sample_pne}
\begin{tabular}{lc}
\hline
Lines used    &  No. of PNe  \\
\hline
{\ariv} $\lambda\lambda$7170+7262        & 23  \\
{\ariv} $\lambda$7262 &   23    \\
{\ariv} $\lambda\lambda$7170+7237+7262 & 9 \\
{\ariv} $\lambda$7170 &  2 \\
Total & 57 \\
\hline
\end{tabular}    
\end{center}
\end{table}

In this work, we selected the PNe subset in DESIRED-E where both {\tel}({\ariv}) and {\tel}({\oiii}) could be computed simultaneously. In DESIRED-E, {\tel}({\ariv}) is derived using the ratio of one or more of the temperature-sensitive auroral {\ariv} $\lambda\lambda$7170,7237,7262 lines to the nebular {\ariv} $\lambda\lambda$4711,4740 lines. We further restricted the sample to objects where {\nel}({\ariv}) could be derived from the {\ariv} $I$($\lambda$4711)/$I$($\lambda$4740) ratio, thereby avoiding observations potentially affected by the blending of {\ariv} $\lambda$4711 with He~{\sc i} $\lambda$4713. Interestingly, by applying this criterion most of the PNe with {\tel}({\ariv}) $>$30,000\,K (all but two), are effectively discarded. 

While such extreme temperatures would theoretically imply an unrealistically hot ionizing star, a severe metal deficiency, or significant shock heating, they more likely arise from the misidentification of very faint auroral lines. Specifically, we found that in 17 spectra from the original PNe subset \citep{Kwitter:03, Milingo:02, Milingo:10, Henry:10} the feature detected around $\lambda$7237 was misidentified as [Ar~{\sc iv}] $\lambda$7237.40 line. However, the lack of detection of the {\ariv} auroral lines at $\lambda\lambda$7170, 7262 --which should be intrinsically brighter-- strongly suggests that this feature is heavily contaminated by, if not entirely due to, a blend of the {\cii} $\lambda\lambda$7236.42,7237.17 permitted lines (see Sect.~\ref{sec:lines}). This misidentification leads to an artificial overestimation of \tel({\ariv}) $>$30,000\,K in all these sources. 

The remaining two objects with computed \tel({\ariv})$>$ 30,000\,K are PB6-1 \citep[][\tel({\ariv})$\sim$32,000\,K]{GarciaRojas:09} and PNG\,354.5+03.3 \citep[][\tel({\ariv})$\sim$70,000\,K]{Tan:24}. Although the former is based on the detection of two auroral lines ($\lambda\lambda$7170, 7262) and the latter on $\lambda$7262 alone, we opted to exclude them to maintain a sample free from potential observational biases or extreme physical regimes that deviate from the core population under study.   

Applying these criteria, we obtain a final sample of 57 PNe with reported \tel({\ariv}) and \tel({\oiii}), for which we adopted {\nel}({\ariv}) to robustly compute {\tel}({\ariv}). Table~\ref{tab:sample_pne} summarises the different sets of temperature-sensitive {\ariv} lines used in the determination of {\tel}({\ariv}). The behaviour of the computed {\tel}({\ariv}) on the availability of the different auroral lines is discussed in Sect.~\ref{sec:results}. 

To compute {\tel}({\oiii}), we followed the standard methodology presented for DESIRED in \citet[][]{MendezDelgado:23b}, using the {\oiii} $\lambda$4363/$\lambda$4959 ratio adopting an average density from all the available density diagnostics. In the presence of very high-density inhomogeneities, this approach may lead to a slight overestimation of {\tel}({\oiii}), which increases with temperature. However, we have tested the effect of adopting the average density or diagnostics sensitive to higher densities, such as {\ariv} or {\feiii} in our sample and found no significant differences in the computed \tel({\oiii}). Therefore, we decided to retain the original density criterion to compute \tel({\oiii}). 

To estimate the errors in the derived diagnostics, uncertainties in the line fluxes were adopted from the original sources and then propagated to the computed physical conditions using 100 Monte Carlo simulations, which yields a reasonable compromise between computing time and convergence of the results. The final adopted values and uncertainties in the physical conditions are the mean and the standard deviation of the 100 individual points, weighted by the inverse square of the error on each quantity.

The computed \tel({\oiii}) and \tel({\ariv}), as well as the average density from all available diagnostics, \nel({\ariv}), and \nel({\feiii}) (see Sect.~\ref{sec:density}) for our PN sample are presented in Table~\ref{table:physical_conditions} in Appendix~\ref{sec:appendix_a}. In Sect.~\ref{sec:tests}, we discuss the sensitivity of the computed \tel({\ariv}) to the selection of the available {\ariv} auroral lines, and to the atomic data sets available in {\sc PyNeb}. 

\section{The \tel({\ariv}) anomaly}
\label{sec:results}

\begin{figure*}[ht!]     
    \centering
     \includegraphics[width=0.80\textwidth]{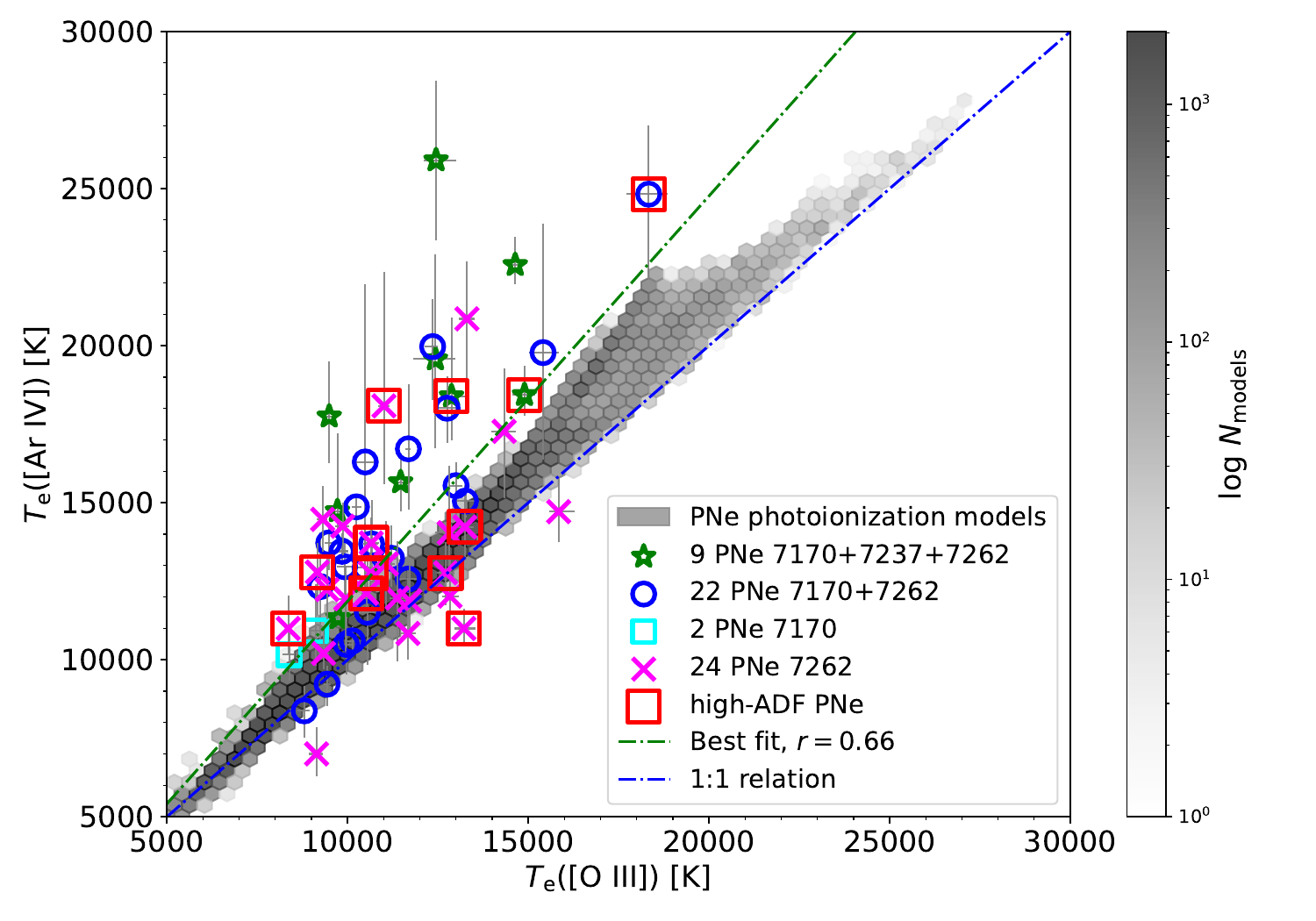}
    \caption{{\tel}({\oiii}) vs. {\tel}({\ariv}) derived by adopting {\nel}({\ariv} $\lambda$4711/$\lambda$4740) to compute {\tel}({\ariv}).  We use different symbols to represent PNe with {\tel}({\ariv}) values computed from various sets of available auroral lines (see Table~\ref{tab:sample_pne}). Open green stars denote PNe with available {\ariv} $\lambda\lambda$7170, 7237, 7262 lines; open blue circles are those with {\ariv} $\lambda\lambda$7170, and 7262 lines; open cyan squares represent cases where only {\ariv} $\lambda$7170 line was measured; and magenta crosses indicate those where only {\ariv} $\lambda$7262 line was measured. PNe with ADF~$>$~5 are marked with slightly bigger open red squares. The dot-dashed blue line shows the 1:1 relation, while the black dashed line indicates a linear fit to the data. Hexagons represent a density plot of photoionisation models in the {\tel}(Ar$^{3+}$) -- {\tel}(O$^{2+}$) plane (see text).}
    \label{fig:Te_ArIV_OIII}
\end{figure*}

In Fig.~\ref{fig:Te_ArIV_OIII} we plot the observed \tel({\ariv}) against \tel({\oiii}) for our PN sample. We use different symbols to represent PNe with $T_{\mathrm{e}}$([Ar~\textsc{iv}]) values computed from various sets of available auroral lines (see Table~\ref{tab:sample_pne}). A detailed inspection of this figure reveals no systematic offsets or differences in the distribution of the data points based on the specific set of auroral lines used to compute \tel({\ariv}). However, since the {\ariv} $\lambda$7237 line is potentially blended with the {\cii} $\lambda\lambda$7236,7237 lines, we discuss further the possible biases of restricting the computation of \tel({\ariv}) using only one or two auroral lines in Sect.~\ref{sec:lines}. On the other hand, it is well established that, in highly excited PNe, with large abundance discrepancy factors (ADFs), recombination can make a non-negligible contribution to the flux of the auroral {\oiii} $\lambda$4363 line \citep[][]{Liu:00, GomezLlanos:20b}. Since a robust correction for this contribution is difficult to obtain due to the lack of corresponding atomic data and the difficult estimation of \tel\ (and O$^{3+}$/O) in this region, the abundance analysis module used in DESIRED-E does not currently attempt to account for it. However, we carefully verified that the behaviour of PNe with ADF $\leq$ 5 is indistinguishable from that of PNe with ADF $>$ 5 in the \tel({\ariv}) versus \tel({\oiii}) diagram (see Fig.~\ref{fig:Te_ArIV_OIII}).

Fig.~\ref{fig:Te_ArIV_OIII} also includes results from photoionisation models in the Mexican Million Models Database\footnote{\href{https://sites.google.com/site/mexicanmillionmodels/}{https://sites.google.com/site/mexicanmillionmodels/}} \citep[3MdB,][]{Morisset:15}, which were computed using Cloudy 17.01 \citep{Ferland:17}. We selected models from the `PNe\_2020' and `PNe\_2021' subsets of the updated 3MdB\_17 database, which span a relatively wide range of metallicities (from log(O/H)=$-$5.465 to $-$2.76). To focus on realistic nebular conditions, we applied the same selection criteria as in \citet{DelgadoInglada:14}, including constraints on stellar luminosity--$T_{\rm eff}$ combinations, ionised hydrogen mass ($<$ 1 M$_{\odot}$), reasonable $n_{\rm H} R_{\rm out}^3$\footnote{\citet{DelgadoInglada:14} excluded the models having at the same time large (small) outer radius and high (low) electron densities, by restricting their trimmed grid of models to 2 $\times 10^{53} \leq n_{\rm H} R_{\rm out}^3 \leq 3 \times 10^{56}$.}, surface brightness, and other physical parameters, which led to a total number of 160\,366 models. The results of these photoionisation models are shown in Fig.~\ref{fig:Te_ArIV_OIII} as a hexagonal binning plot, representing the density of models (indicated by the color bar) in the $T_{\rm e}$(Ar$^{3+}$) versus $T_{\rm e}$(O$^{2+}$) plane.  We do not expect large systematic discrepancies between the model-predicted electron temperatures (computed directly from the thermal structure of the ionised gas) and those inferred from observational diagnostics, since both are based on consistent atomic data and assume thermal equilibrium conditions. In the models, temperatures are ionic fraction-weighted averages using:
\begin{equation}
T_{\rm e}(X^i) = \frac{\int_V T_{\rm e}\cdot n_{\rm e}\cdot n_{\rm H}\cdot ff\cdot (X^i/X)\cdot dV}{\int_V n_{\rm e}\cdot n_{\rm H}\cdot ff\cdot (X^i/X)\cdot dV},
\end{equation}

\noindent where $T_{rm e}$ and $n_{\rm e}$ are the electron temperature and density, $n_{\rm H}$ is the hydrogen density, $ff$ is the filling factor, and X$^{i}$/X is the ionic fraction. The values of $T_{\rm e}$(X$^i$) are obtained from the {\it teion} table of 3MdB. These values provide a close analogue to observationally derived diagnostic temperatures. Although slight differences may arise due to spatial variations and weighting effects, a direct comparison remains meaningful. In practice, for the figures presented in this paper, we associate the observed emission line ratio temperature with the theoretical ionic fraction-weighted temperature, for example, T$_{\rm e}$(O$^{++}$) and T$_{\rm e}$ ([OIII]).

\begin{figure*}[ht!]
    \centering
    \begin{tabular}{cc}
        \includegraphics[width=0.5\textwidth]{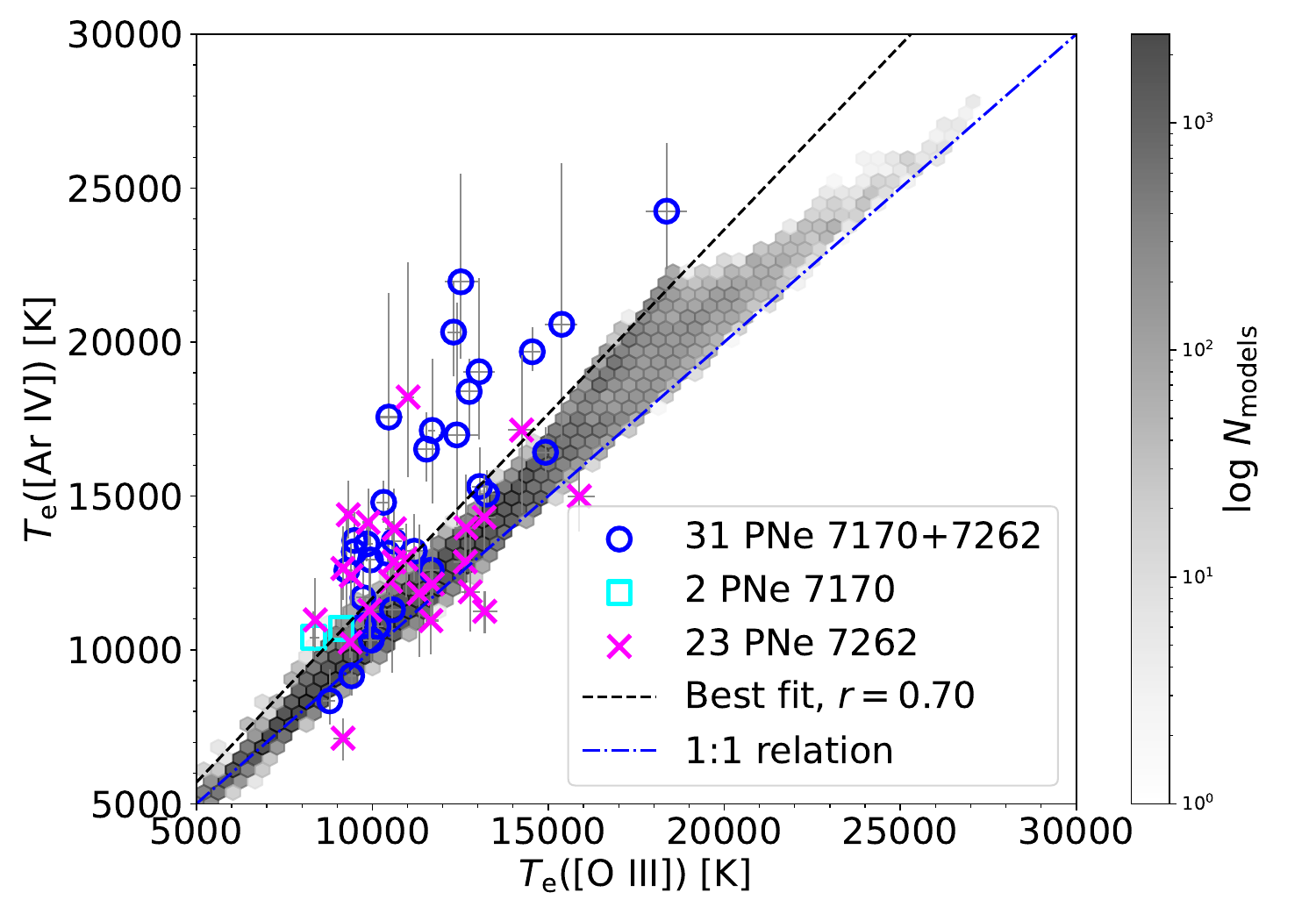} &
        \includegraphics[width=0.5\textwidth]{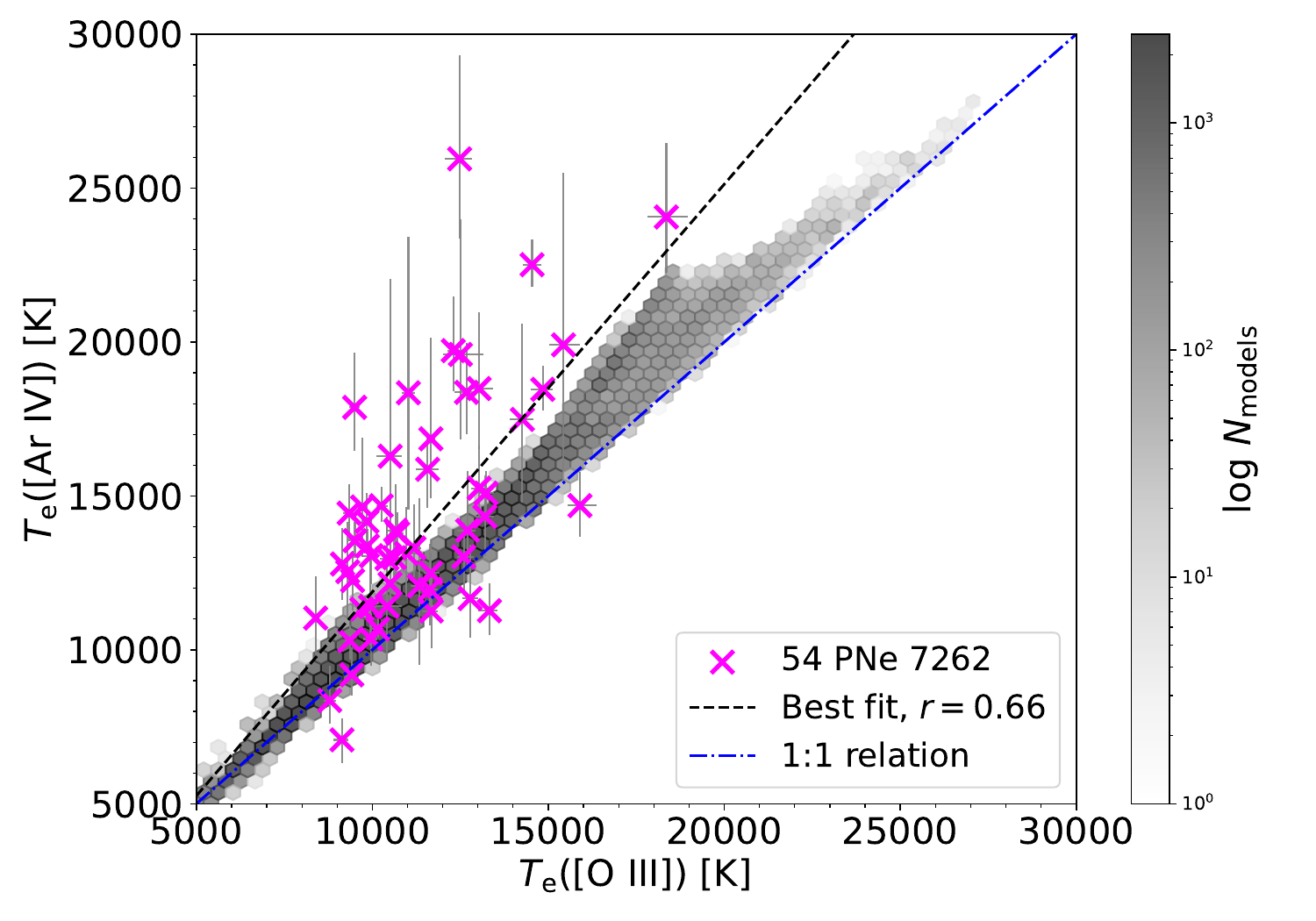} 
    \end{tabular}
    \caption{Comparison of electron temperature relations derived from different [Ar~{\sc iv}] auroral line selections. The left panel shows $T_e$([O~{\sc iii}]) versus $T_e$([Ar~{\sc iv}]) excluding the [Ar~{\sc iv}] $\lambda$7237 line. The right panel shows the same relation using only the [Ar~{\sc iv}] $\lambda$7262 line. Selection criteria are identical to those used in Fig.~\ref{fig:Te_ArIV_OIII}.}
    \label{fig:auroral_lines}
\end{figure*}

A visual inspection of Fig.~\ref{fig:Te_ArIV_OIII}, reveals that the observational data behave differently from photoionisation models predictions: in many cases, {\tel}({\ariv}) is considerably higher than expected from the models. A two-sample Kolmogorov-Smirnov (KS) test comparing the observed \tel({\ariv}) distribution to the model grid yields a statistic of $0.27$ with $p < 0.001$. This result indicates that the observations are statistically incompatible with the distributions predicted by the photoionization models. The failure of the photoionisation models to reproduce the high-temperature tail of the observed distribution suggests that the {\ariv} temperature excess is a real feature of the data. However, before discussing possible physical interpretations, it is essential to evaluate the impact of observational biases and methodological uncertainties on the derived temperatures.

\section{Robustness tests and observational constraints}
\label{sec:tests}

\subsection{Effect of auroral line selection}
\label{sec:lines}

To discard possible effects on the selection of the auroral lines to compute \tel({\ariv}) we made a careful revision of the reported fluxes of the auroral {\ariv} lines in the DESIRED-E database.

The Ar$^{3+}$ ion has a $p^3$ ground-state configuration, which is analogous to that of O$^{+}$. Consequently, as we have already pointed out in Sect.~\ref{sec:desired}, it gives rise to several auroral transitions: [Ar~{\sc iv}] $\lambda$7170.62 ($^2$P$_{3/2}$--$^2$D$_{3/2}$), $\lambda$7262.76 ($^2$P$_{1/2}$--$^2$D$_{3/2}$), $\lambda$7237.40 ($^2$P$_{3/2}$--$^2$D$_{5/2}$), and $\lambda$7331.40 ($^2$P$_{1/2}$--$^2$D$_{5/2}$), with relative ratios normalized to $\lambda$7170.62 of 1:0.85:0.74:0.16 (at \tel=10,000~K and \nel=1000 cm$^{-3}$), respectively. The latter is at least a factor of five fainter than the other components and is located in close proximity to the [O~{\sc ii}] $\lambda\lambda$7320, 7330 auroral doublet, making its detection particularly challenging. 
While the remaining three lines are expected to have comparable fluxes, their detection is highly sensitive to the radial velocity of the source, as they are frequently affected by telluric emission and absorption features or line blending. In particular, [Ar~{\sc iv}] $\lambda$7237.40 can be blended with the C~{\sc ii} $\lambda\lambda$7236.42, 7237.17 lines, which would artificially increase the measured flux and, consequently, the derived {\tel}({\ariv}), or even saturate the diagnostic \citep[see e.g.][]{Keenan:97}. To test the robustness of our results against such effects, we re-evaluated the temperature relation using different subsets of these lines. We applied a slightly different strategy than \citet{Keenan:97}, who investigated the behaviour of \tel({\ariv}) using each auroral line individually.

 \begin{figure*}[ht!]
    \centering
    \begin{tabular}{cc}
        \includegraphics[width=0.5\textwidth]{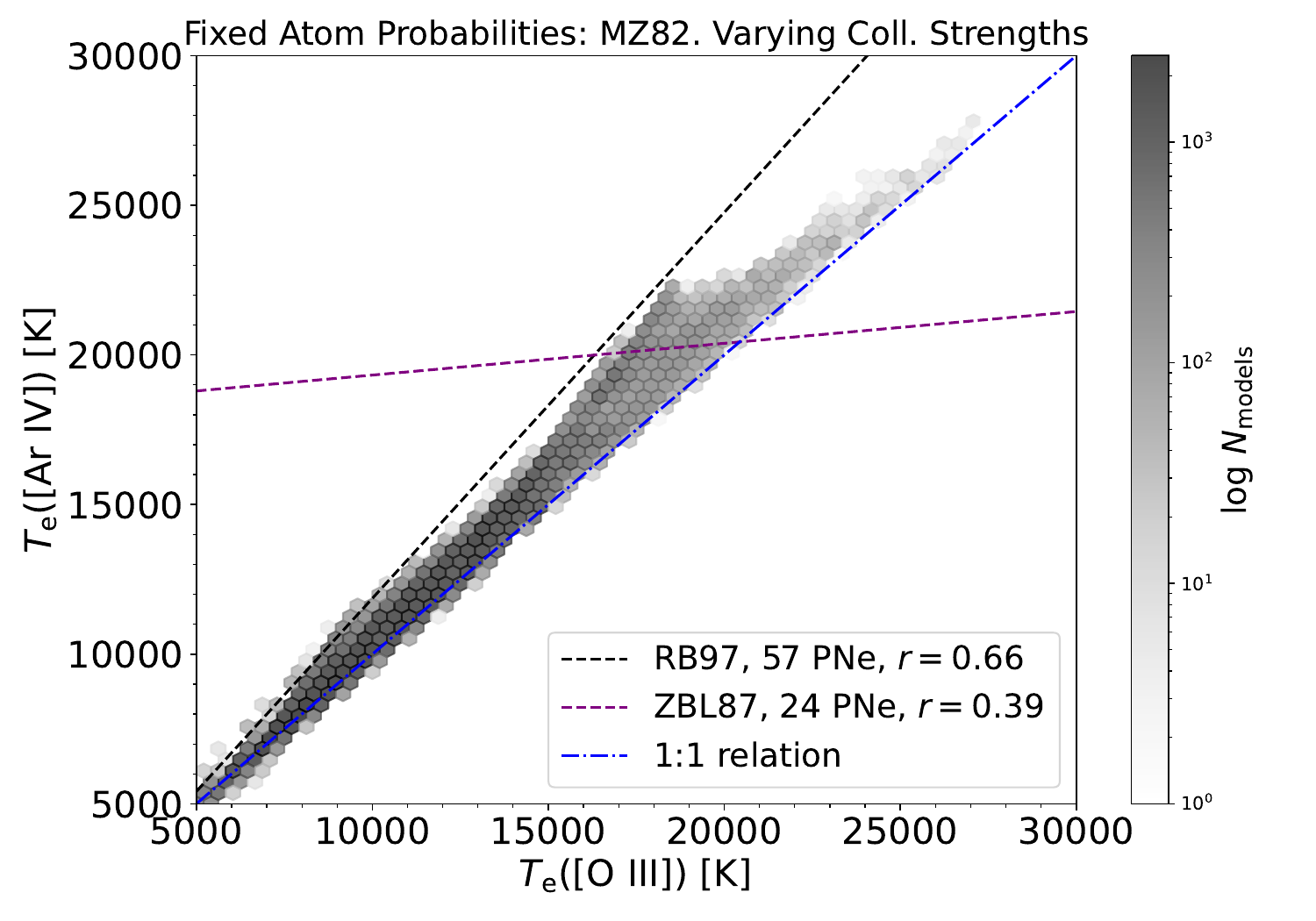} &
        \includegraphics[width=0.5\textwidth]{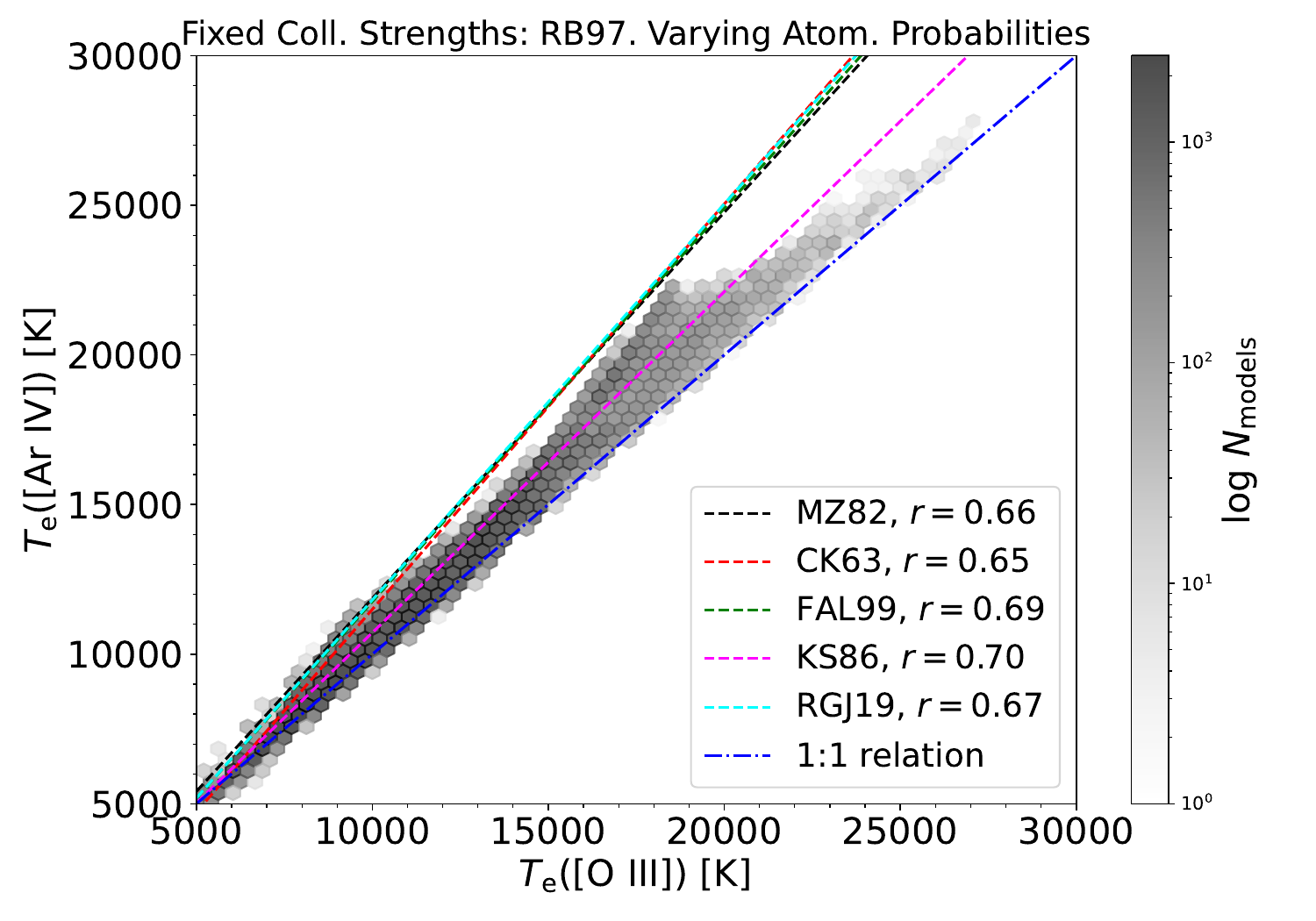} 
    \end{tabular}
    \caption{Comparison of the fit to the relation between {\tel}({\ariv}) and {\tel}({\oiii}) using different atomic data sets available in {\sc PyNeb}. The default datasets are those of \citet[][MZ82]{Mendoza:82a} for transition probabilities and \citet[][RB97]{Ramsbottom:97} for collision strengths. In both panels the fit shown in Fig.~\ref{fig:Te_ArIV_OIII} using the default atomic data is represented as a black dashed line, and the 1:1 relation as a dashed-dotted blue line. Left panel: Fixed default atomic probabilities and varying collision strengths. Right panel: Fixed default collision strengths and varying transition probabilities. Similar to Fig.~\ref{fig:Te_ArIV_OIII}, in each panel, hexagonal binning plots display the density of photoionisation models in the corresponding temperature–temperature plane. References and acronyms are defines in Table~\ref{tab:atomic_data}. All teh fits in these panel include the whole set of 57 PNe.}
    \label{fig:Atom_Coll_comp}
\end{figure*}

In Fig.~\ref{fig:auroral_lines}, we investigate how the choice of auroral lines impacts the calculation of $T_e$([Ar~{\sc iv}]). The left panel displays the $T_e$([O~{\sc iii}]) vs. $T_e$([Ar~{\sc iv}]) relation—initially presented in Fig.~\ref{fig:Te_ArIV_OIII}—but omitting the contribution of the [Ar~{\sc iv}] $\lambda$7237 line. The right panel shows the results obtained using only the [Ar~{\sc iv}] $\lambda$7262 line. In both instances, we applied the same selection criteria as in Fig.~\ref{fig:Te_ArIV_OIII}. It is evident that an excess of objects with high $T_e$([Ar~{\sc iv}]) persists in both cases, although the linear fits show slightly better agreement with the predictions from photoionisation models. Notably, the PNe with $T_e$([Ar~{\sc iv}]) $> 30,000$~K that were previously excluded by the $n_e$([Ar~{\sc iv}]) criterion were also filtered out here without the need of applying the \nel criterion. This reinforces the conclusion that deep, high-spectral resolution data are essential for deriving reliable $T_e$([Ar~{\sc iv}]) values.

\subsection{Dependence on atomic data}
\label{sec:atomic}

\begin{table*}
    \centering    
    \caption{References for atomic data used for Ar$^{3+}$, O$^{2+}$, and Fe$^{2+}$.}
    \label{tab:atomic_data}
    \begin{tabular}{lcc}
        \hline
        \noalign{\smallskip}
        Ion & Transition probabilities & Collision strengths\\
        \noalign{\smallskip}
        \hline
        \noalign{\smallskip}
            O$^{2+}$  &  \citet{Wiese:96}, \citet{Storey:00} & \citet{Storey:14}\\
            Fe$^{2+}$  &  \citet{Quinet:96} & \citet{Zhang:96}\\
            Ar$^{3+}$ &   \citet{Mendoza:82a}  & \citet{Ramsbottom:97}\\
                      &   \citet{Czyzak:63}  & \citet{Mendoza:83b}\\
                      &   \citet{Mendoza:82a}, \citet{Kaufman:86}  & \citet{Zeippen:87}\\
                      &   \citet{Fritzsche:99}  &   \\
                      &   \citet{Rynkun:19}  &   \\
        \noalign{\smallskip}
        \hline
    \end{tabular}
\end{table*}

Following a similar approach to that considered by \citet{JuandeDios:17}, we have studied the influence of adopting different sets of atomic data on the behaviour of \tel([Ar~{\sc iv}]). The available atomic data in {\sc PyNeb} for the ions studied in this work are summarised in Table~\ref{tab:atomic_data}. 

The default atomic data set in {\sc PyNeb} v1.1.30 consists of the collision strengths from \citet{Ramsbottom:97} and the transition probabilities from \citet{Mendoza:82a}. We combined different datasets by fixing either the default transition probabilities or collision strengths and varying the other atomic data with the sets available in {\sc PyNeb}. We then recomputed both \nel({\ariv}) and \tel({\ariv}), and determined the best fit for the \tel({\oiii}) vs. \tel({\ariv}) relation in each case. The collision strengths of \citet{Mendoza:83b} were not considered, following the recommendations by \citet{JuandeDios:17}, as they imply densities differing significantly from other data sets available for the same diagnostic.

Figure~\ref{fig:Atom_Coll_comp} illustrates the fits derived from this methodology, presented alongside the original fit from Fig.~\ref{fig:Te_ArIV_OIII} (dashed black line), the 1:1 relation (dash-dotted blue line), and the same suite of photoionisation models shown in Fig.~\ref{fig:Te_ArIV_OIII}. An inspection of the left panel of Fig.~\ref{fig:Atom_Coll_comp} reveals that collision strengths from \citet{Zeippen:87} provide a poorer fit to the data. The resulting sample consists of only 23 PNe because this set of collision strengths was only tabulated up to \tel=20,000~K\footnote{\citet{Zeippen:87} also provided collision strengths for \tel=50,000~K, but labeled them as highly unreliable due to the omission of higher resonances.}, yielding {\tt NaN} values for ratios corresponding to temperatures outside of this validity range. Moreover, these set of collision strengths makes the relation exhibits a slope significantly flatter than that predicted by photoionization models. This behaviour suggests that this specific atomic data set may not be suitable for this diagnostic. In the right panel of Fig.~\ref{fig:Atom_Coll_comp}, it is clear that varying the transition probability datasets does not modify the general behaviour of the {\tel}--{\tel} relation. However, the combination of transition probabilities of \citet{Mendoza:82a} and \citet{Kaufman:86} appears to provide better agreement with the general trend predicted by the photoionisation models. Nevertheless, the discrepancy remains robust: approximately 31\% of the PNe sample exhibits a \tel({\ariv}) more than $2\sigma$ higher than photoionization model predictions. This confirms that the offset is not merely an artifact of the adopted atomic data, as the distribution remains systematically shifted even when accounting for measurement errors. That said, caution is warranted as only two sets of collision strengths are currently available in {\sc PyNeb}; a different temperature dependence in the collision strengths could still potentially account for this discrepancy.

\begin{figure}[ht!]     
    \centering
    {\includegraphics[width=0.5\textwidth]{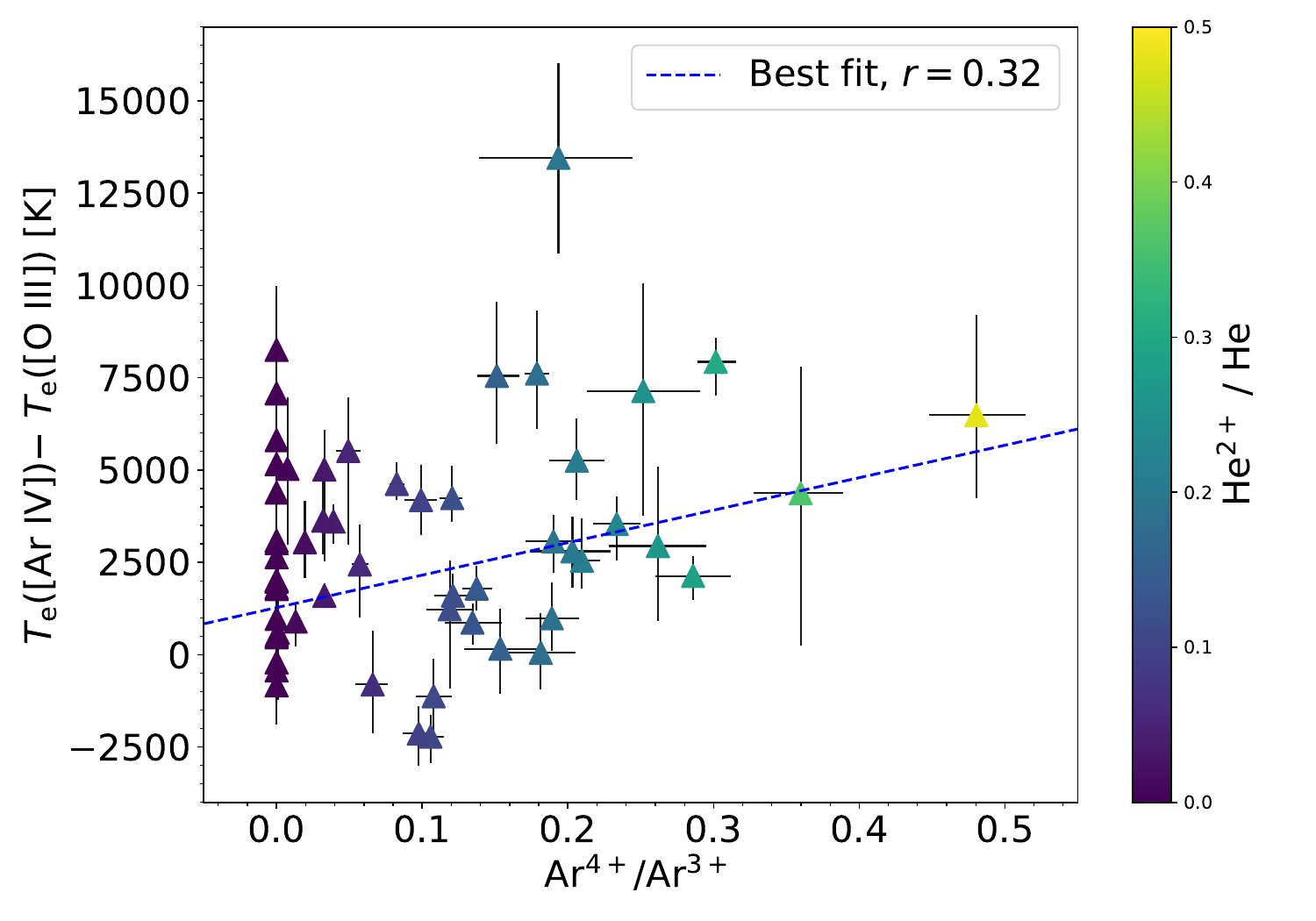}}
    \caption{Difference between \tel({\ariv}) and \tel({\oiii}) as a function of the $\text{Ar}^{4+}/\text{Ar}^{3+}$ ratio. The color bar indicates the nebular excitation based on the observed $\text{He}^{2+}/\text{He}$ ratio. An Orthogonal Distance Regression (ODR) fit, excluding objects with $\text{Ar}^{4+}/\text{Ar}^{3+} = 0$, is shown as a dashed blue line.}
    \label{fig:Diff_Ar_ioniz}
\end{figure}

Finally, given that the \ariv-emitting zone is located within the $\text{He}^{2+}$ region --where $\text{Ar}^{4+}$ can be abundant-- and that auroral transitions are energetically favorable recombination cascades, a recombination contribution to the auroral \ariv\ lines emerges as a plausible explanation for the observed discrepancies in \tel({\ariv}). Although recombination coefficients for \ariv\ are currently unavailable, we examined the behavior of \tel({\ariv}) in relation to PN excitation. Figure~\ref{fig:Diff_Ar_ioniz} shows the difference between \tel({\ariv}) and \tel({\oiii}) versus the $\text{Ar}^{4+}/\text{Ar}^{3+}$ ratio; objects where \arv\ lines were not detected are set to $\text{Ar}^{4+}/\text{Ar}^{3+} = 0$ and are excluded from the fit. Although a slight correlation is present, it lacks statistical significance ($r = 0.32$), preventing any definitive conclusion regarding the impact of recombination on the observed \tel({\ariv}) behaviour.

\subsection{Possible density biases}
\label{sec:density}

\begin{figure}[ht!]     
    \centering
    {\includegraphics[width=0.5\textwidth]{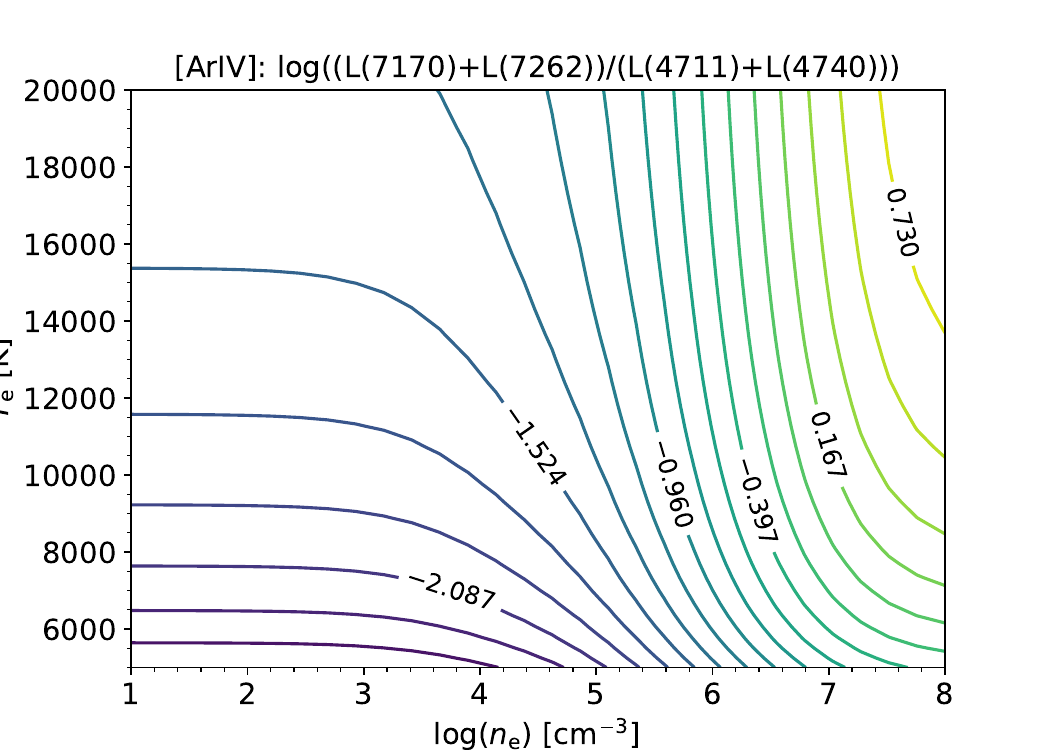}}
    \caption{Predicted dependence of the \tel-sensitive {\ariv} ($\lambda\lambda$7170+7262)/($\lambda\lambda$4711+4740) intensity ratios with physical conditions. This ratio progressively changes from being \tel-sensitive to \nel-sensitive in the range (\nel\ $\sim 10^3-10^5$ cm$^{-3}$), becoming purely \nel-sensitive at nebular temperatures for densities $> 10^5$ cm$^{-3}$. }
    \label{fig:Ar4_Diag}
\end{figure}

In the first study using DESIRED spectra of \hii\ regions, \citet{MendezDelgado:23b} found that \nel of the objects can be underestimated when {\sii} and {\oii} diagnostics are the only available density indicators, leading to an overestimation of \tel({\sii}) and \tel({\oii}). In our case, the sensitivity of the \tel({\ariv}) diagnostic on the density becomes critical for values above 10$^5$ cm$^{-3}$, as illustrated in Fig.~\ref{fig:Ar4_Diag}. In the hypothetical presence of high-density clumps with \nel  $\gtrsim 10^5$ cm$^{-3}$, the {\ariv} density diagnostic is almost insensitive to the presence of such clumps; consequently, adopting a density lower than the actual value would lead to an overestimation of \tel({\ariv}). 

\begin{figure}[ht!]     
    \centering
    {\includegraphics[width=0.5\textwidth]{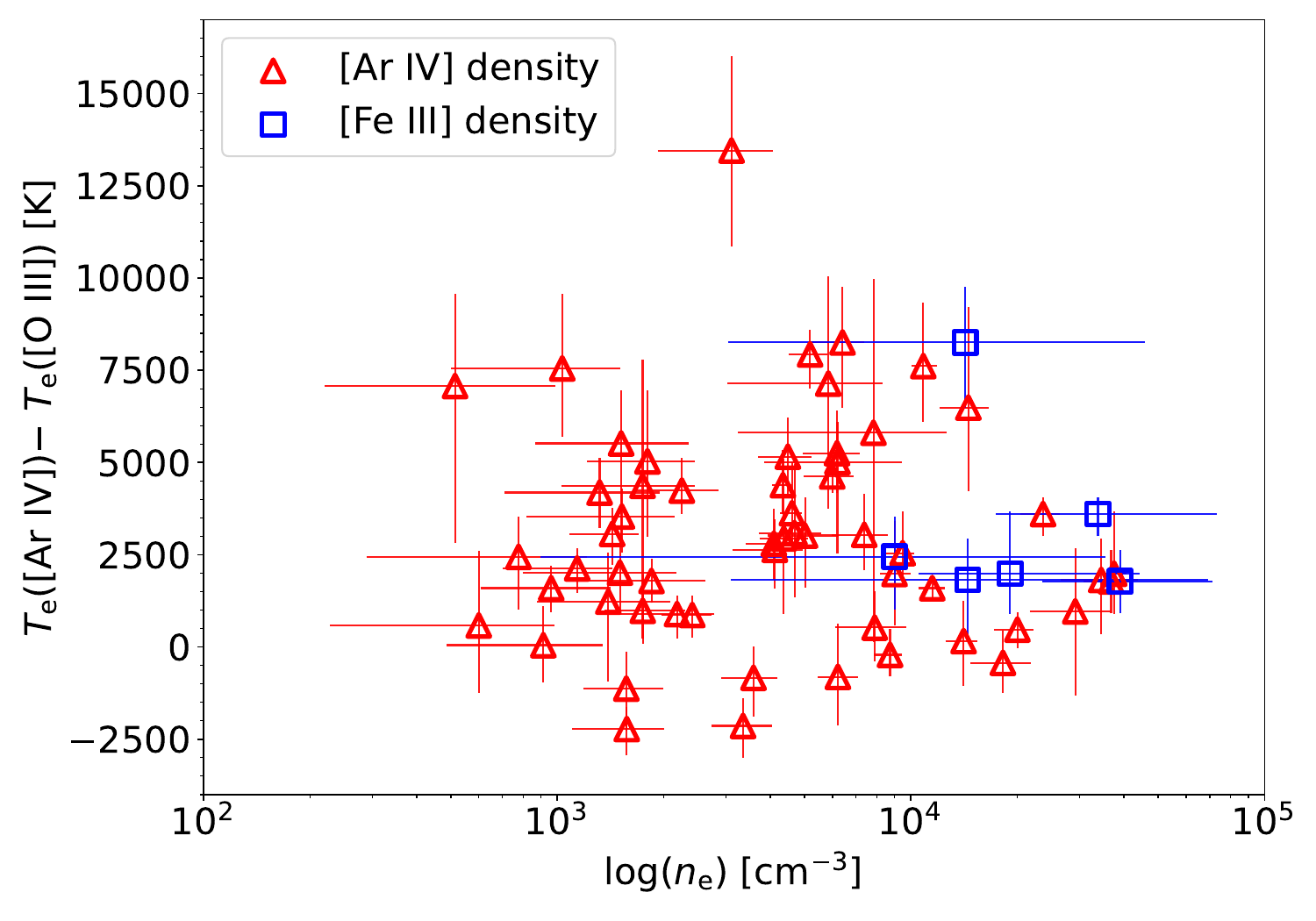}}
    \caption{Difference \tel({\ariv})$-$\tel({\oiii}) versus the computed density using {\ariv} (red triangles) or {\feiii} (blue squares) diagnostic ratios for our PN sample. }
    \label{fig:deltaT}
\end{figure}

As a first step to investigate the possible effect of density inhomogeneities in our PNe sample, we plot the temperature difference, \tel({\ariv})$-$\tel({\oiii}) against \nel({\ariv}) in Fig.\ref{fig:deltaT} (red triangles). No dependence of the magnitude of the temperature anomaly with the density is found. To further verify the presence of high-density inhomogeneities, we examined the behaviour of this temperature difference relative to \nel({\feiii} $\lambda$4658/$\lambda$4701) (blue squares) in Fig.~\ref{fig:Ar4_Diag}, which remains sensitive to density up to 10$^6$ cm$^{-3}$ \citep{MendezDelgado:23b}. The atomic data adopted for the computations of \nel({\feiii}) are summarised in Table~\ref{tab:atomic_data}. Although this diagnostic could only be computed for six objects and, in principle would provide densities in a zone where Ar$^{3+}$ does not coexist, no significant differences in the temperature difference relative to the density were found. 

It is worth noting that one could always find a specific density value that forces consistency between the \tel({\ariv}) and \tel({\oiii}) temperatures. However, this would require a density significantly higher than the one derived from {\ariv} --an outcome that would imply that the high-ionisation zone is orders of magnitude denser than the low-ionisation regions (represented by {\sii} or {\oii}), which is physically counter-intuitive, given the observed anti-correlation between density and degree of ionisation. In summary, if the temperature anomaly were driven by a severe underestimation of \nel, such an effect would remain intrinsically hidden, as the {\ariv} diagnostic itself becomes insensitive in this high-density regime. This would necessitate the existence of extremely dense clumps ($\gtrsim 10^5$~cm$^{-3}$) within the high-ionisation gas that remain undetected by standard optical indicators—a possibility that, while difficult to rule out entirely, finds no support in our current multi-diagnostic analysis.

\subsection{Excitation diagnostics: testing for shocks} 
\label{sec:shocks}

\begin{figure*}[ht!]     
    \centering
    {\includegraphics[width=\textwidth]{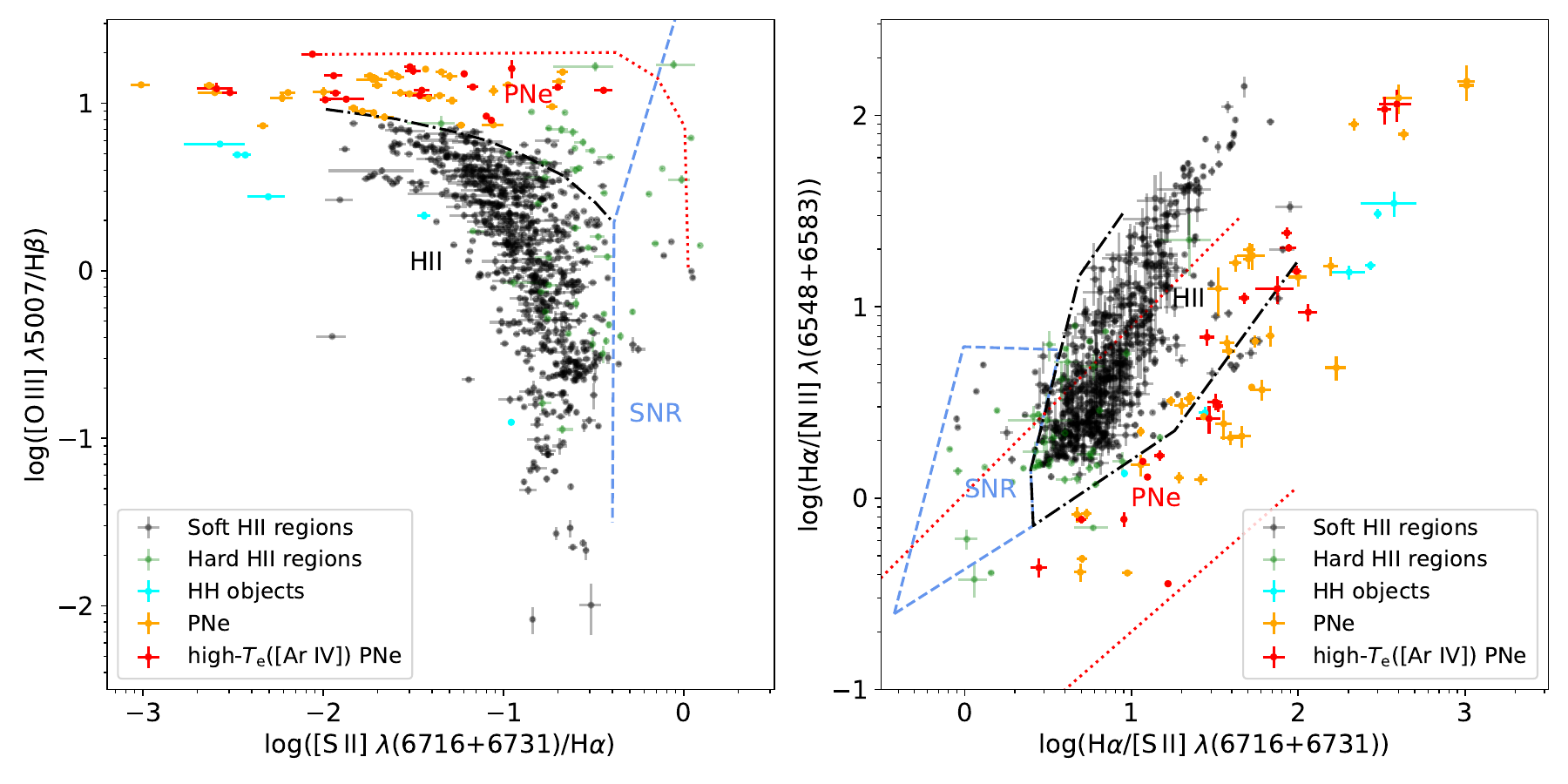}}
    \caption{Left panel: Excitation diagrams log({\oiii} $\lambda5007$/H$\beta$) versus log({\sii} $\lambda\lambda6716+6731$/H$\alpha$) proposed by \citet{Sabin:13}. Right panel: Excitation diagram log(H$\alpha$/({\nii} $\lambda\lambda6548+6583$)) versus log(H$\alpha$/({\sii} $\lambda\lambda6716+66731$)) proposed by \citet{Sabbadin:77}. In both cases, the limits are the ones proposed by \citet{Sabin:13}. Black dots correspond to ``soft'' Galactic and extragalactic {\hii} regions in DESIRED-E that are below the empirical relation by \citet{Kauffmann:03} in the BPT diagram. Green dots correspond to ``hard'' {\hii} regions in DESIRED-E, i.e. they are above  the empirical relation by \citet{Kauffmann:03} in the BPT diagram. Orange dots represent the 57 PNe in our DESIRED-E sample, and cyan dots display photoionised HH objects. Red dots correspond to PNe that are above the linear fit in the left panel of Fig.~\ref{fig:Te_ArIV_OIII}, i.e., they show a higher {\tel}({\ariv}) than predicted by photoionisation models.}
    \label{fig:SMB77_diagram}
\end{figure*}

To explore the possible influence of other excitation mechanisms beyond pure photoionisation on the behaviour of  $T_{\rm e}$([Ar~{\sc iv}]), we plotted our data on some excitation diagnostic diagrams available in the literature. These diagrams are specifically designed to differentiate between different types of nebulae and excitation conditions. The results are presented in Fig.~\ref{fig:SMB77_diagram}. The left panel shows log({\oiii} $\lambda5007$/H$\beta$) versus log({\sii} $\lambda\lambda6716+6731$/H$\alpha$) \citep{Sabin:13}, while the right panel shows log(H$\alpha$/({\nii} $\lambda\lambda6548+6583$)) versus log(H$\alpha$/({\sii} $\lambda\lambda6716+6731$)) \citep{Sabbadin:77}, with the boundaries defined by \citet{Sabin:13}. In both diagrams, we overplot our sample of 57 PNe (orange and red dots), together with data for Galactic and extragalactic {\hii} regions from the DESIRED-E database. The {\hii} regions have been labelled as ``soft'' (black dots) or ``hard'' (green dots) depending on whether they lie below or above the empirical relation by \citet{Kauffmann:03} in the Baldwin-Phillips-Terlevich (BPT) diagram \citep{Baldwin:81}, respectively \citep[see Fig.~1 in][]{Garciarojas:25}. We also include some photoionised Herbig-Haro (HH) objects from DESIRED-E (cyan dots). These spectra were obtained in the bow shocks of HH objects in the Orion nebula that have been photoionised after the passage of the shock front, owing to the intense radiation field in the nebula \citep[see e.g.,][]{MendezDelgado:21a,MendezDelgado:21b} although they still show some signs of the effect of the passage of a shock.

To test whether PNe with high \tel({\ariv}) exhibit signs of shock excitation, we highlighted in red those PNe for which \tel{\ariv} deviates more than 20\% from the temperature obtained adopting the best-fit relation shown in Fig.~\ref{fig:Te_ArIV_OIII}. However, these PNe do not show a clear trend or preference for the shock-dominated regions of the Sabbadin diagrams. While some objects appear near the area occupied by HH objects, the majority remain within the expected locus for photoionised PNe. This lack of a clear correlation suggests that, while shocks may be present in some cases, they are unlikely to be the primary driver of the systematically high temperatures observed across the sample. Instead, the discrepancy may be rooted in the complex thermal structure of the high-ionisation gas.

\subsection{Comparison with other high-ionisation diagnostics}
\label{sec:otherdiags}

\begin{figure*}[htbp]
    \centering
    \begin{tabular}{cc}
        \includegraphics[width=0.5\textwidth]{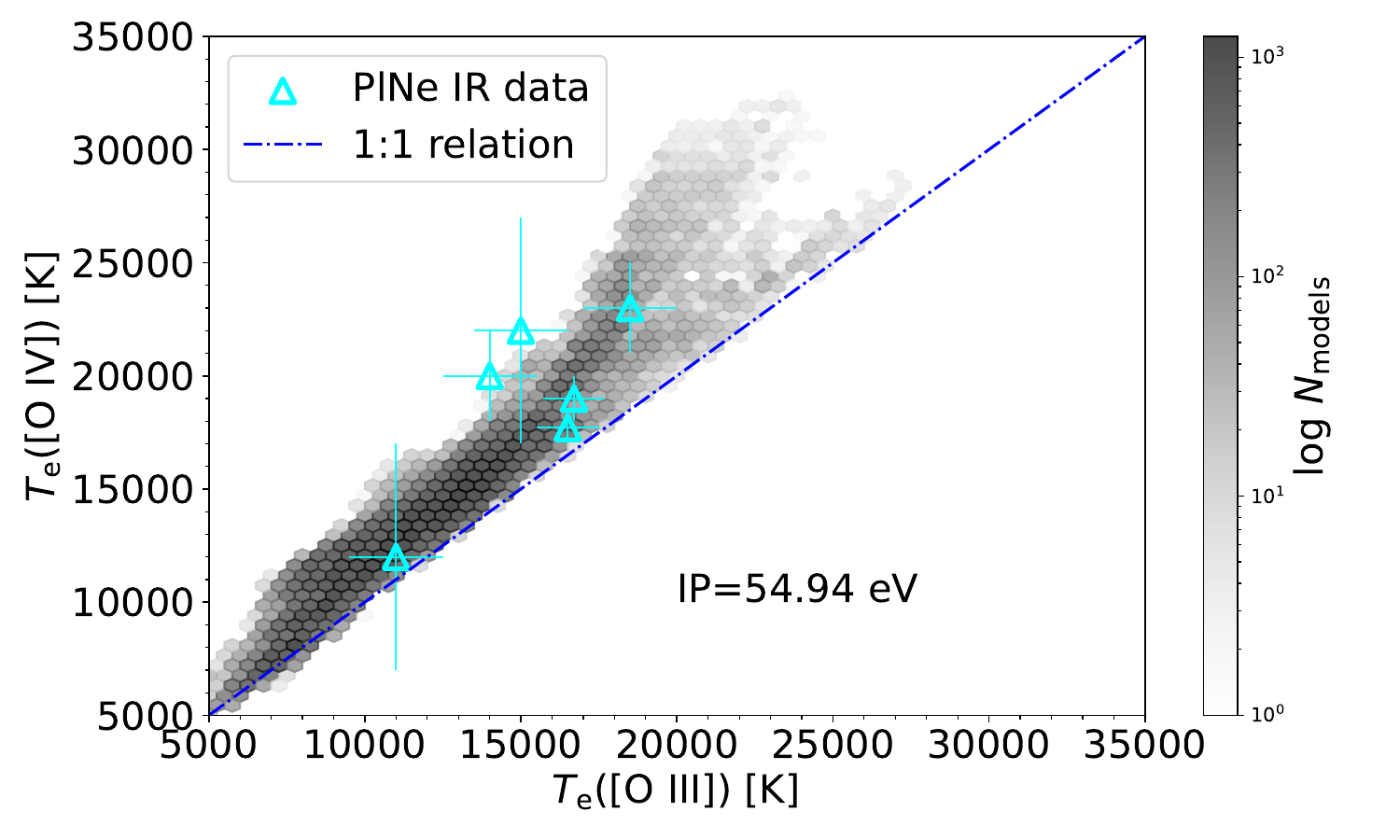} &
        \includegraphics[width=0.5\textwidth]{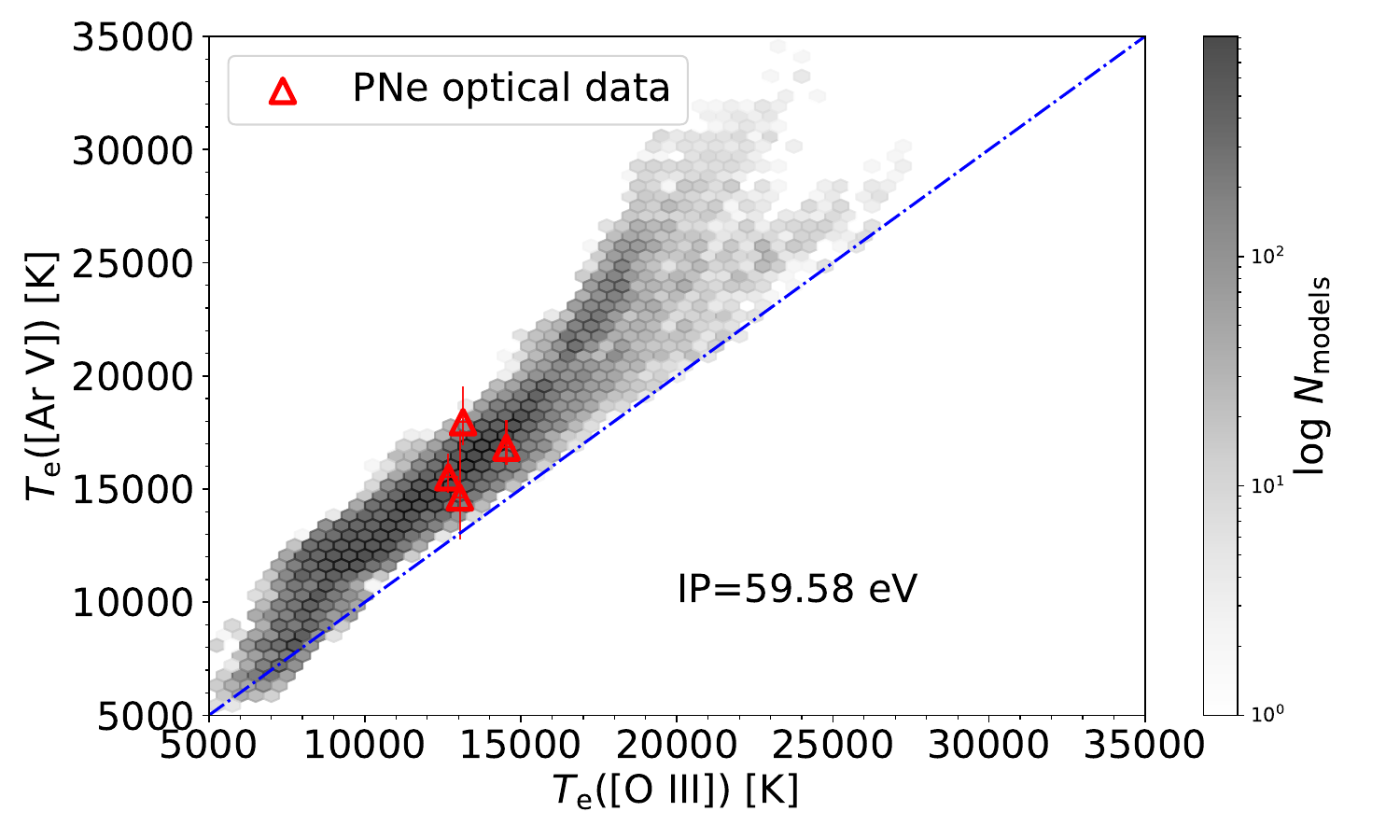} \\

        \includegraphics[width=0.5\textwidth]{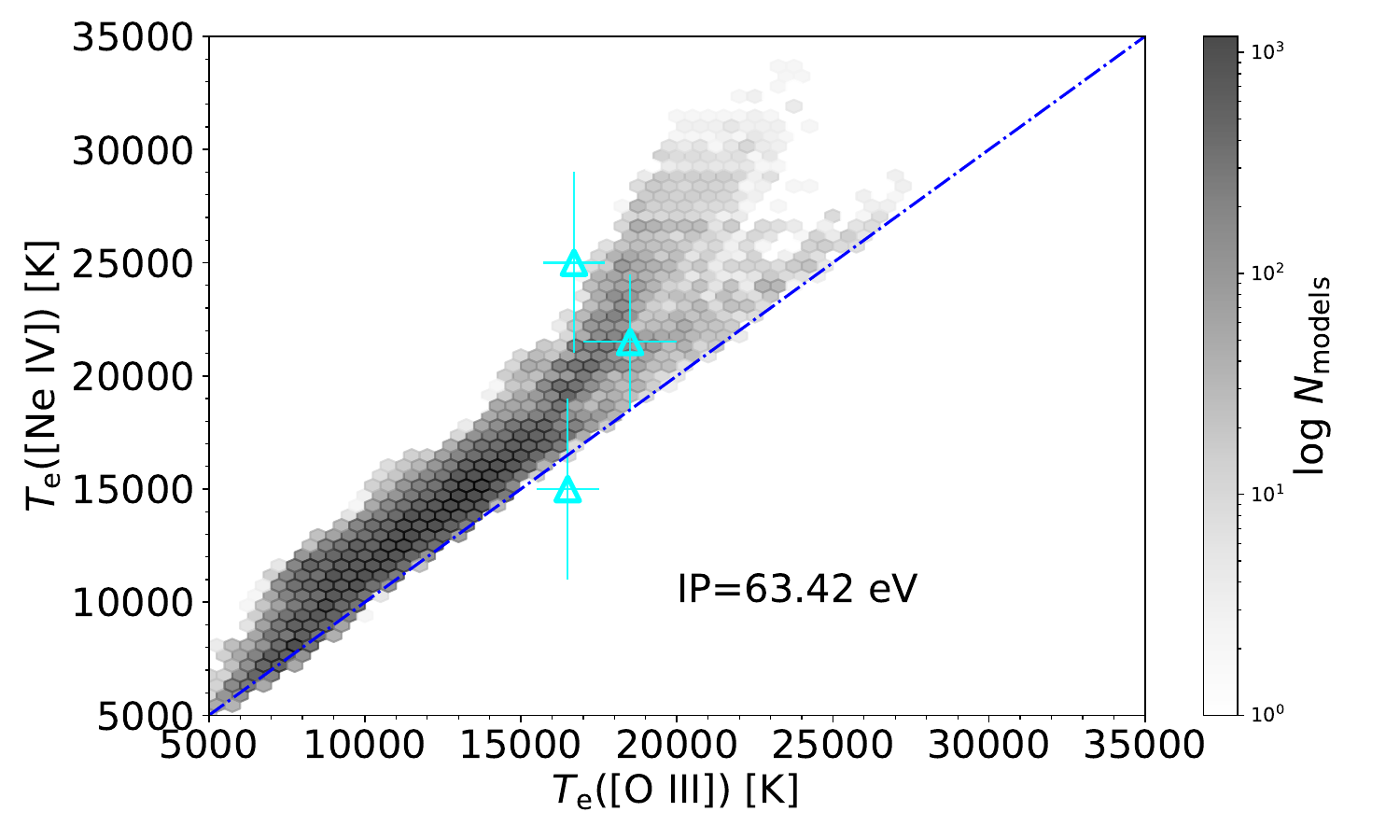} &
        \includegraphics[width=0.5\textwidth]{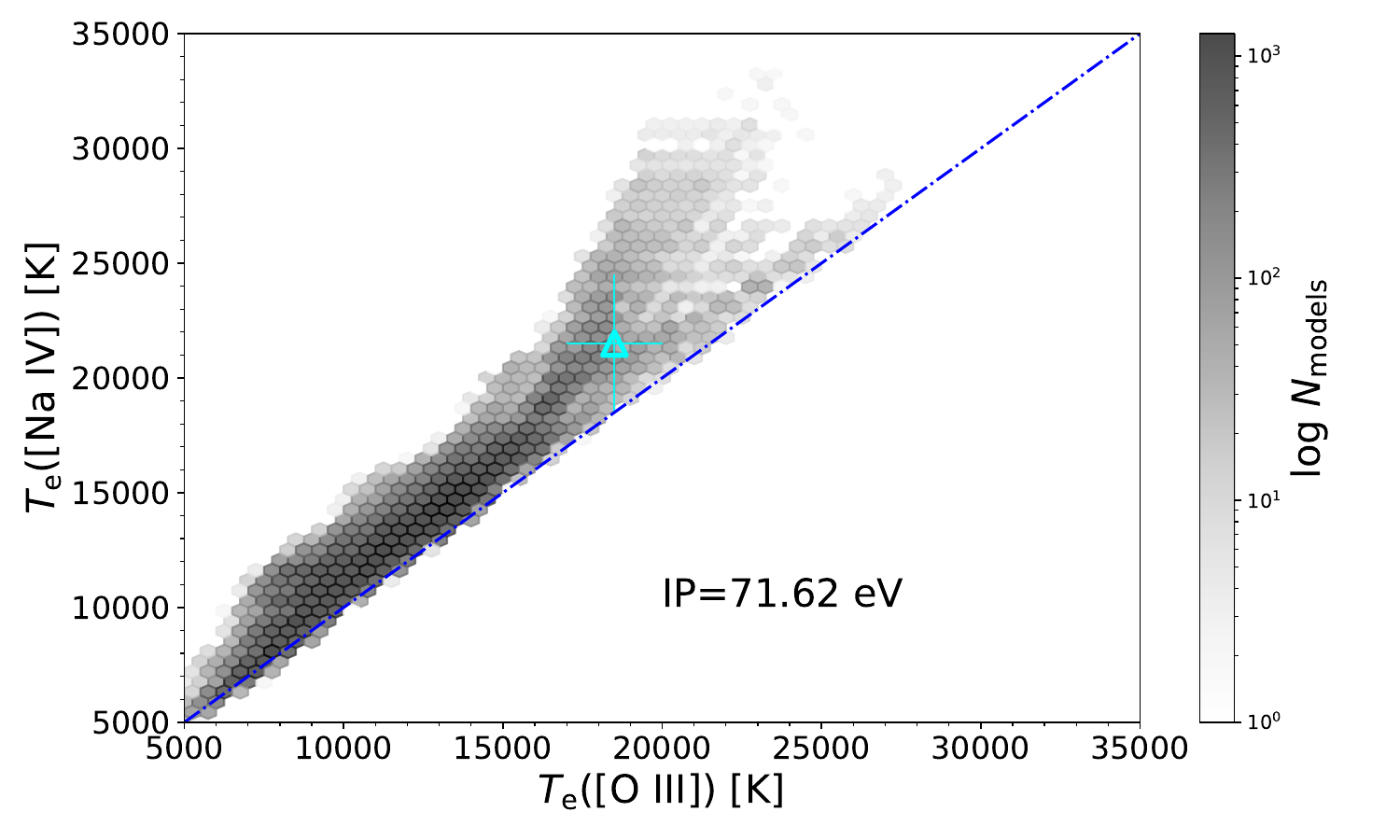} \\

        \includegraphics[width=0.5\textwidth]{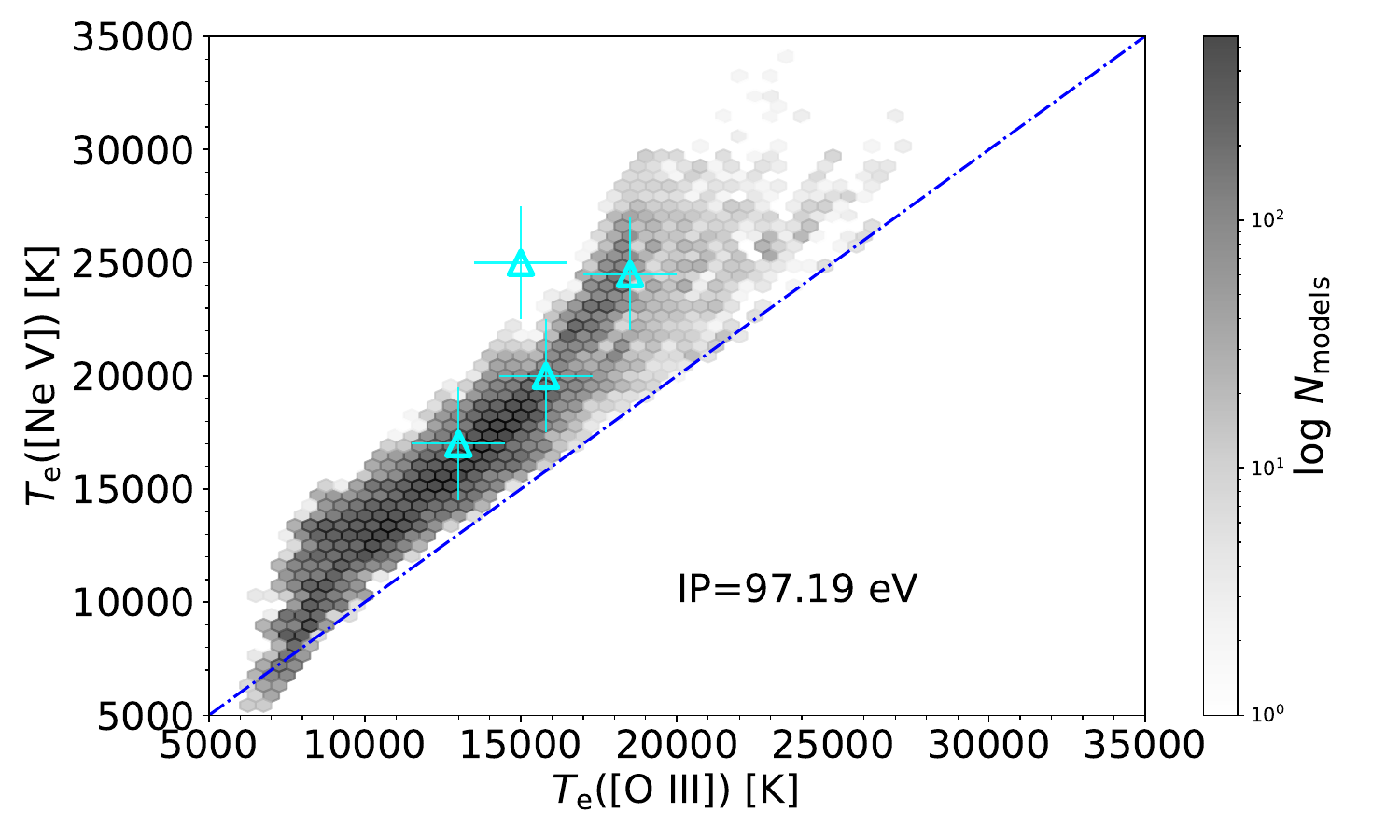} &
        \includegraphics[width=0.5\textwidth]{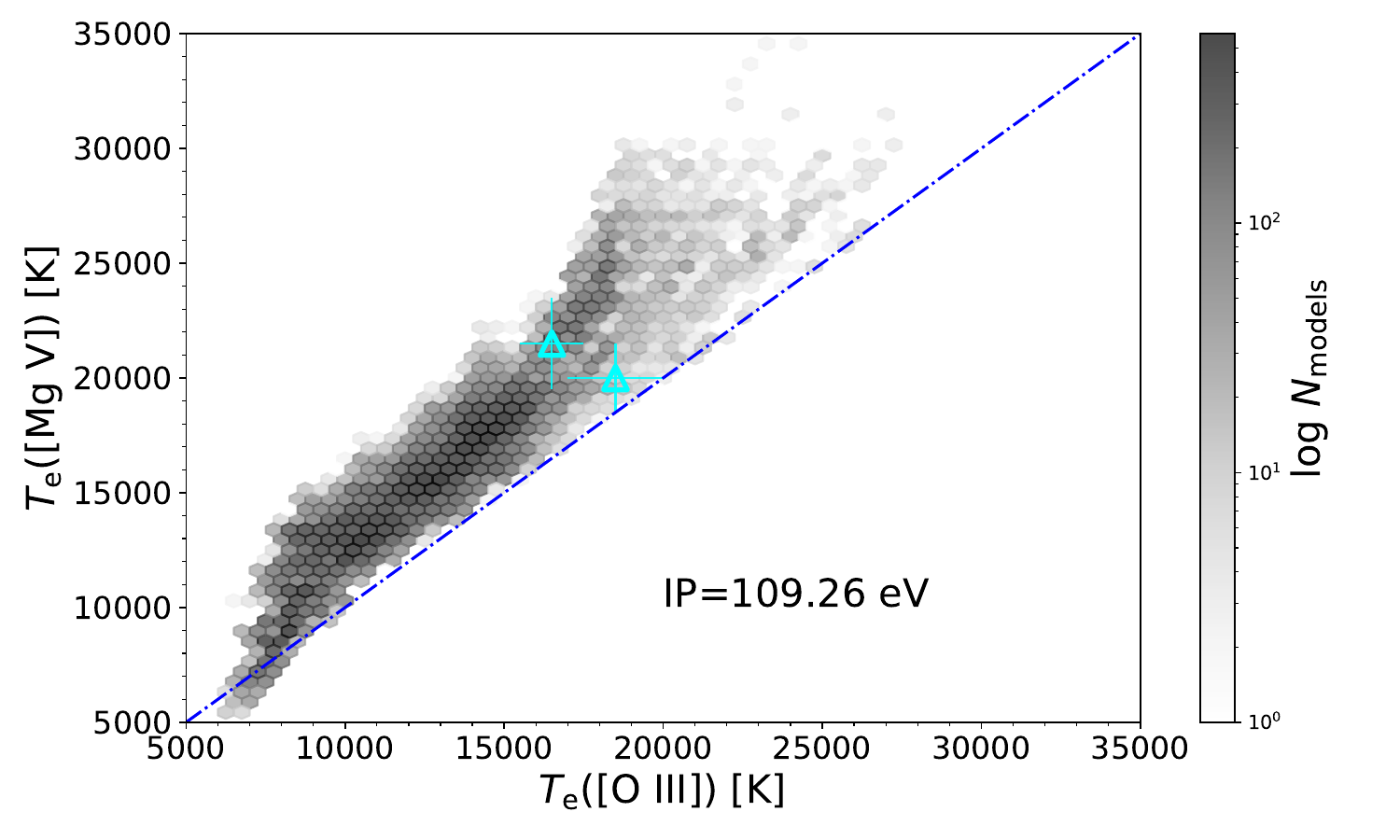}
    \end{tabular}
    \caption{Relation between different temperature diagnostics and \tel({\oiii}). Panels are arranged according to the ionisation potential of the ion shown on the y-axis. In each panel, hexagonal binning plots display the density of photoionisation models in the corresponding temperature–temperature plane (see text). Only models with an ionic fraction ($X^{+i}/X$) of the higher-ionisation species larger than 5\% were included. Red triangles represent PNe for which optical diagnostic ratios could be measured. Cyan triangles indicate PNe with \tel\ derived from infrared diagnostics reported in the literature (see text). The dashed line on each panel shows the 1:1 relation.}
    \label{fig:Te_OIII_others}
\end{figure*}

This complexity is further evidenced when examining other high-ionisation species. In a series of studies devoted to a detailed analysis of individual high-excitation PNe, \citet{Pottasch:99, Pottasch:00, Pottasch:07, Pottasch:08a, Pottasch:08b}, and \citet{Pottasch:09} combined UV, optical, and IR data to determine the physical conditions in the very high-ionisation regime. These authors employed diagnostics based on O~{\sc iv}] (IP$\sim$55 eV), [Ne~{\sc iv}] (IP$\sim$63 eV), [Na~{\sc iv}] (IP$\sim$72 eV), [Ne~{\sc v}] (IP$\sim$97 eV), and [Mg~{\sc v}] (IP$\sim$109 eV).
Taking advantage of the high sensitivity of the DESIRED-E spectra, we explored the behaviour of additional high-ionisation temperature diagnostics, such as the {\arv} $\lambda$4625/$\lambda$7005 (IP$\sim$60 eV) ratio. Fig.~\ref{fig:Te_OIII_others} displays the relations between these various \tel\ diagnostics and \tel({\oiii}) from our optical data and from IR literature data. The panels are arranged  from top to bottom and from left to right according to increasing ionisation potential. In all cases, 3MdB photoionisation models are overplotted as hexagonal binning maps, representing the density of models in the corresponding temperature–temperature planes. We verified that the model distributions are robust by  restricting the sample to models where the ionic fractions of the high-ionisation species in question exceeds 5\%, thereby eliminating spurious temperature determinations.
The observed behaviour in all panels of Fig.~\ref{fig:Te_OIII_others} is that while the measured temperatures are high, the 3MdB photoionisation models also predict high values for these ions. In those cases, the observations and models are largely in agreement. The discrepancy is therefore not a global ``high-ionisation temperature excess'', but appears to be specifically localized to {\tel}({\ariv}). However, we must acknowledge the limited statistics in the cases shown in Fig~\ref{fig:Te_OIII_others}, where the maximum number of PNe available for comparison is only six in the case of \tel({\oiv}).

Although none of the individual tests provides a definitive explanation for the observed {\ariv} temperature excess, the combined evidence indicates that the effect is robust and likely linked to the specific physical conditions of the Ar$^{3+}$-emitting region rather than to general properties of the ionised gas. 

\section{On the origin of the {\ariv} temperature anomaly}
\label{sec:discuss}

The comparison with other high-ionisation diagnostics shows that the temperature excess identified in {\tel}({\ariv}) is not part of a general temperature enhancement affecting the innermost nebular regions. In contrast to the behaviour of {\ariv}, the temperatures derived from ions probing comparable or even higher ionisation potentials are broadly consistent with the predictions of photoionisation models. This indicates that the discrepancy is specific to the Ar$^{3+}$ zone and is therefore likely connected to the particular physical or atomic processes governing this ion.

This selective behaviour places strong constraints on its possible origin. If a general heating mechanism, such as the photoelectric effect on dust grains \citep{Dopita:00, Stasinska:02}, were acting globally in the high-ionisation gas, a systematic shift would be expected in all high-ionisation temperature diagnostics. The absence of such a trend argues against a general heating effect and instead points to processes associated specifically with the Ar$^{3+}$ ionisation layer or with the atomic physics underlying the {\ariv} diagnostic.

As discussed in Sect.~\ref{sec:intro}, Ar$^{3+}$ occupies a narrow transition region between the O$^{2+}$ and He$^{2+}$ zones, where the ionisation structure and thermal balance are expected to vary rapidly. This may make the {\ariv} temperature diagnostic particularly sensitive to departures from the simplified thermal structure assumed in standard photoionisation models or to uncertainties in the available atomic data, especially in the temperature dependence of the collision strengths. Improving the interpretation of this diagnostic will require both more complete atomic calculations and tailored photoionisation models capable of resolving the thermal structure of this transition region. 

\section{Summary}
\label{sec:conclu}

In this work, we have performed a homogeneous analysis of the {\ariv} electron temperature in a large sample of PNe from the DESIRED-E database. Our main conclusions are as follows:

\begin{enumerate}
    \item We find a systematic and localized ``temperature excess'' in {\tel}({\ariv}) compared to {\tel}({\oiii}). Unlike other high-ionisation species whose high temperatures are well-reproduced by 3MdB models, {\tel}({\ariv}) consistently exceeds model expectations.
    \item The observed discrepancy is robust against observational and systematic artifacts, including the choice of auroral lines ($\lambda$7170, $\lambda$7237, $\lambda$7262) and the variety of atomic data sets currently available in {\sc PyNeb}. However, we note that since only two sets of collision strengths are available for {\ariv}, a potential temperature dependence in future atomic calculations could still play a role in resolving this offset.
    \item Excitation diagnostic diagrams indicate that shock excitation is not a universal explanation for this behaviour, as the PNe with the largest {\tel}({\ariv}) offsets do not coincide with the shock-dominated regions in shock diagnostic plots.
    \item The fact that models successfully predict the temperatures of higher-IP ions (e.g., [Ne~{\sc v}], [Mg~{\sc v}]) but fail for {\ariv} suggests that the discrepancy is not due to a global inner-nebula heating mechanism like dust. Instead, it points toward a more localized physical effect or a potential limitation in our understanding of the ionisation structure and atomic data specific to the Argon ion stages.
\end{enumerate}

These results identify {\tel}({\ariv}) as a uniquely sensitive (and currently problematic) diagnostic of the physical conditions in PNe. Accurately resolving this anomaly is essential for the reliable determination of chemical abundances in high-excitation nebulae.

\section*{Acknowledgements}
We thank the anonymous referee for their time and comments on the manuscript. We want to thank S. Tan for providing the extinction corrected fluxes of FORS2 spectra of bulge PNe. JGR, CE and MOG acknowledge financial support from the Agencia Estatal de Investigaci\'on of the Ministerio de Ciencia e Innovaci\'on y Universidades (AEI- MCIU) under grant ``The internal structure of ionised nebulae and its effects in the determination of the chemical composition of the interstellar medium and the Universe'' with reference PID2023-151648NB-I00 (DOI:10.13039/501100011033). JGR and DJ acknowledge support from the AEI-MCIU and from the European Regional Development Fund (ERDF) under grant ``Planetary nebulae as the key to understanding binary stellar evolution''  with reference PID2022-136653NA-I00 (DOI:10.13039/501100011033). 
CM acknowledges the support of UNAM/DGAPA/PAPIIT grants IG101223 and IN113226.

\section*{DATA AVAILABILITY}

All data used in this study are publicly available in the original publications from which the emission-line fluxes and/or intensities were compiled. Appendix~\ref{sec:appendix_a} provides the complete list of references and computed physical conditions corresponding to the spectroscopic data comprising the DESIRED-E subsample of PNe presented in this work.
Software: 
\texttt{PyNeb} \citep{Luridiana:15, Morisset:20, Mendoza:23}, 
\texttt{numpy} \citep{numpy}, 
\texttt{scipy} \citep{scipy}.
\texttt{pandas} \citep{pandas:10,pandas:20},
\texttt{matplotlib} \citep{matplotlib},
\texttt{jupyter} \citep{jupyter}.

\bibliographystyle{aa}
\bibliography{refs} 

@ARTICLE{MendezDelgado:23b,
author = {{M{\'e}ndez-Delgado}, J.~E. and {Esteban}, C. and {Garc{\'\i}a-Rojas}, J. and {Arellano-C{\'o}rdova}, K.~Z. and {Kreckel}, K. and {G{\'o}mez-Llanos}, V. and {Egorov}, O.~V. and {Peimbert}, M. and {Orte-Garc{\'\i}a}, M.},
title = "{Density biases and temperature relations for DESIRED H II regions}",
journal = {\mnras},
year = 2023,
month = aug,
volume = {523},
number = {2},
pages = {2952-2973},
doi = {10.1093/mnras/stad1569},
archivePrefix = {arXiv},
eprint = {2305.13136},
primaryClass = {astro-ph.GA},
adsurl = {https://ui.adsabs.harvard.edu/abs/2023MNRAS.523.2952M}
}

@ARTICLE{Peimbert:67,
author = {{Peimbert}, Manuel},
title = "{Temperature Determinations of H II Regions}",
journal = {\apj},
year = 1967,
month = dec,
volume = {150},
pages = {825},
doi = {10.1086/149385},
adsurl = {https://ui.adsabs.harvard.edu/abs/1967ApJ...150..825P}
}

@ARTICLE{Peimbert:17,
author = {{Peimbert}, Manuel and {Peimbert}, Antonio and {Delgado-Inglada}, Gloria},
title = "{Nebular Spectroscopy: A Guide on Hii Regions and Planetary Nebulae}",
journal = {\pasp},
year = 2017,
month = aug,
volume = {129},
number = {978},
pages = {082001},
doi = {10.1088/1538-3873/aa72c3},
archivePrefix = {arXiv},
eprint = {1705.06323},
primaryClass = {astro-ph.GA},
adsurl = {https://ui.adsabs.harvard.edu/abs/2017PASP..129h2001P}
}

@ARTICLE{Ferland:17,
author = {{Ferland}, G.~J. and {Chatzikos}, M. and {Guzm{\'a}n}, F. and {Lykins}, M.~L. and {van Hoof}, P.~A.~M. and {Williams}, R.~J.~R. and {Abel}, N.~P. and {Badnell}, N.~R. and {Keenan}, F.~P. and {Porter}, R.~L. and {Stancil}, P.~C.},
title = "{The 2017 Release Cloudy}",
journal = {\rmxaa},
year = 2017,
month = oct,
volume = {53},
pages = {385-438},
doi = {10.48550/arXiv.1705.10877},
archivePrefix = {arXiv},
eprint = {1705.10877},
primaryClass = {astro-ph.GA},
adsurl = {https://ui.adsabs.harvard.edu/abs/2017RMxAA..53..385F}
}

@Article{Wiese:96,
author = {{Wiese}, W.~L. and {Fuhr}, J.~R. and {Deters}, T.~M.},
title = "{Atomic transition probabilities of carbon, nitrogen, and oxygen : a critical data compilation}",
journal = {Journal of Physical and Chemical Reference Data, Monograph 7},
booktitle = {Atomic transition probabilities of carbon, nitrogen, and oxygen : a critical data compilation.~ Edited by W.L.~Wiese, J.R.~Fuhr, and T.M.~Deters.~Washington, DC :  American Chemical Society ...~for the National Institute of Standards and Technology (NIST) c1996.~QC 453 .W53 1996.~ Also Journal of Physical and Chemical Reference Data, Monograph 7.~ Melville, NY: AIP Press},
year = 1996,
volume = 403,
adsurl = {https://ui.adsabs.harvard.edu/abs/1996atpc.book.....W}
}

@ARTICLE{Storey:00,
author = {{Storey}, P.~J. and {Zeippen}, C.~J.},
title = "{Theoretical values for the [Oiii] 5007/4959 line-intensity ratio and homologous cases}",
journal = {MNRAS},
year = 2000,
month = mar,
volume = 312,
pages = {813-816},
doi = {10.1046/j.1365-8711.2000.03184.x},
adsurl = {http://adsabs.harvard.edu/abs/2000MNRAS.312..813S}
}

@ARTICLE{Storey:14,
author = {{Storey}, P.~J. and {Sochi}, T. and {Badnell}, N.~R.},
title = "{Collision strengths for nebular [O III] optical and infrared lines}",
journal = {\mnras},
year = 2014,
month = jul,
volume = 441,
pages = {3028-3039},
doi = {10.1093/mnras/stu777},
adsurl = {http://adsabs.harvard.edu/abs/2014MNRAS.441.3028S}
}

@ARTICLE{Fritzsche:99,
author = {{Fritzsche}, S. and {Fricke}, B. and {Geschke}, D. and {Heitmann}, A. and
{Sienkiewicz}, J.~E.},
title = "{Forbidden Transitions in the Ground-State Configuration of Low-Z Phosphorus-like Ions}",
journal = {\apj},
year = "1999",
month = "Jun",
volume = {518},
number = {2},
pages = {994-1001},
doi = {10.1086/307328},
adsurl = {https://ui.adsabs.harvard.edu/abs/1999ApJ...518..994F}
}

@INPROCEEDINGS{Mendoza:83b,
author = {{Mendoza}, C.},
title = "{Recent advances in atomic calculations and experiments of interest in the study of planetary nebulae}",
booktitle = {Planetary Nebulae},
year = "1983",
editor = {{Aller}, L.~H.},
series = {IAU Symposium},
volume = {103},
month = "Jan",
pages = {143-172},
adsurl = {https://ui.adsabs.harvard.edu/abs/1983IAUS..103..143M}
}

@ARTICLE{Kaufman:86,
author = {{Kaufman}, Victor and {Sugar}, Jack},
title = "{Forbidden Lines in ns$^{2}$np$^{k}$ Ground Configurations and nsnp Excited Configurations of Beryllium through Molybdenum Atoms and Ions}",
journal = {Journal of Physical and Chemical Reference Data},
year = "1986",
month = "Jan",
volume = {15},
number = {1},
pages = {321-426},
doi = {10.1063/1.555775},
adsurl = {https://ui.adsabs.harvard.edu/abs/1986JPCRD..15..321K}
}

@ARTICLE{Ramsbottom:97,
author = {{Ramsbottom}, C.~A. and {Bell}, K.~L.},
title = "{Effective Collision Strengths for Electron-Impact Excitation of Triphy Ionized Argon}",
journal = {Atomic Data and Nuclear Data Tables},
year = "1997",
month = "Jan",
volume = {66},
pages = {65},
doi = {10.1006/adnd.1997.0741},
adsurl = {https://ui.adsabs.harvard.edu/abs/1997ADNDT..66...65R}
}

@ARTICLE{Quinet:96,
author = {{Quinet}, P.},
title = "{Transition probabilities for forbidden lines of Fe III.}",
journal = {\aaps},
year = "1996",
month = "May",
volume = {116},
pages = {573-578},
adsurl = {https://ui.adsabs.harvard.edu/abs/1996A&AS..116..573Q}
}

@ARTICLE{Zhang:96,
author = {{Zhang}, H.},
title = "{Atomic data from the Iron Project. XVIII. Electron impact excitation collision strengths and rate coefficients for Fe III.}",
journal = {\aaps},
year = 1996,
month = nov,
volume = {119},
pages = {523-528},
adsurl = {https://ui.adsabs.harvard.edu/abs/1996A&AS..119..523Z}
}

@ARTICLE{Tsamis:03,
author = {{Tsamis}, Y.~G. and {Barlow}, M.~J. and {Liu}, X. -W. and {Danziger}, I.~J. and {Storey}, P.~J.},
title = "{A deep survey of heavy element lines in planetary nebulae - I. Observations and forbidden-line densities, temperatures and abundances}",
journal = {\mnras},
year = 2003,
month = oct,
volume = {345},
number = {1},
pages = {186-220},
doi = {10.1046/j.1365-8711.2003.06972.x},
archivePrefix = {arXiv},
eprint = {astro-ph/0305469},
primaryClass = {astro-ph},
adsurl = {https://ui.adsabs.harvard.edu/abs/2003MNRAS.345..186T}
}

@ARTICLE{Baldwin:81,
author = {{Baldwin}, J.~A. and {Phillips}, M.~M. and {Terlevich}, R.},
title = "{Classification parameters for the emission-line spectra of extragalactic objects.}",
journal = {\pasp},
year = 1981,
month = feb,
volume = {93},
pages = {5-19},
doi = {10.1086/130766},
adsurl = {https://ui.adsabs.harvard.edu/abs/1981PASP...93....5B}
}

@ARTICLE{Kauffmann:03,
author = {{Kauffmann}, Guinevere and {Heckman}, Timothy M. and {Tremonti}, Christy and {Brinchmann}, Jarle and {Charlot}, St{\'e}phane and {White}, Simon D.~M. and {Ridgway}, Susan E. and {Brinkmann}, Jon and {Fukugita}, Masataka and {Hall}, Patrick B. and {Ivezi{\'c}}, {\v{Z}}eljko and {Richards}, Gordon T. and {Schneider}, Donald P.},
title = "{The host galaxies of active galactic nuclei}",
journal = {\mnras},
year = 2003,
month = dec,
volume = {346},
number = {4},
pages = {1055-1077},
doi = {10.1111/j.1365-2966.2003.07154.x},
archivePrefix = {arXiv},
eprint = {astro-ph/0304239},
primaryClass = {astro-ph},
adsurl = {https://ui.adsabs.harvard.edu/abs/2003MNRAS.346.1055K}
}

@ARTICLE{MendezDelgado:23a,
author = {{M{\'e}ndez-Delgado}, J. Eduardo and {Esteban}, C{\'e}sar and {Garc{\'\i}a-Rojas}, Jorge and {Kreckel}, Kathryn and {Peimbert}, Manuel},
title = "{Temperature inhomogeneities cause the abundance discrepancy in H II regions}",
journal = {\nat},
year = 2023,
month = jun,
volume = {618},
number = {7964},
pages = {249-251},
doi = {10.1038/s41586-023-05956-2},
archivePrefix = {arXiv},
eprint = {2305.11578},
primaryClass = {astro-ph.GA},
adsurl = {https://ui.adsabs.harvard.edu/abs/2023Natur.618..249M}
}

@ARTICLE{Luridiana:15,
author = {{Luridiana}, V. and {Morisset}, C. and {Shaw}, R.~A.},
title = "{PyNeb: a new tool for analyzing emission lines. I. Code description and validation of results}",
journal = {\aap},
year = 2015,
month = jan,
volume = {573},
eid = {A42},
pages = {A42},
doi = {10.1051/0004-6361/201323152},
archivePrefix = {arXiv},
eprint = {1410.6662},
primaryClass = {astro-ph.IM},
adsurl = {https://ui.adsabs.harvard.edu/abs/2015A&A...573A..42L}
}

@ARTICLE{Mendoza:23,
author = {{Mendoza}, Claudio and {M{\'e}ndez-Delgado}, Jos{\'e} E. and {Bautista}, Manuel and {Garc{\'\i}a-Rojas}, Jorge and {Morisset}, Christophe},
title = "{Atomic Data Assessment with PyNeb: Radiative and Electron Impact Excitation Rates for [Fe II] and [Fe III]}",
journal = {Atoms},
year = 2023,
month = apr,
volume = {11},
number = {4},
eid = {63},
pages = {63},
doi = {10.3390/atoms11040063},
archivePrefix = {arXiv},
eprint = {2304.01298},
primaryClass = {astro-ph.GA},
adsurl = {https://ui.adsabs.harvard.edu/abs/2023Atoms..11...63M}
}

@article{Espiritu:21,
adsurl = {https://ui.adsabs.harvard.edu/abs/2021MNRAS.508.2668E},
archiveprefix = {arXiv},
author = {{Esp{\'\i}ritu}, Jos{\'e} N. and {Peimbert}, Antonio},
doi = {10.1093/mnras/stab2746},
eprint = {2109.10546},
journal = {\mnras},
month = {December},
number = {2},
pages = {2668-2687},
primaryclass = {astro-ph.GA},
title = {{Physical conditions and chemical abundances in PN M 2-36. Results from deep echelle observations}},
volume = {508},
year = {2021}
}

@article{GarciaRojas:12,
adsurl = {https://ui.adsabs.harvard.edu/abs/2012A&A...538A..54G},
archiveprefix = {arXiv},
author = {{Garc{\'\i}a-Rojas}, J. and {Pe{\~n}a}, M. and {Morisset}, C. and {Mesa-Delgado}, A. and {Ruiz}, M.~T.},
doi = {10.1051/0004-6361/201118217},
eid = {A54},
eprint = {1111.4992},
journal = {\aap},
month = {February},
pages = {A54},
primaryclass = {astro-ph.GA},
title = {{Analysis of chemical abundances in planetary nebulae with [WC] central stars. I. Line intensities and physical conditions}},
volume = {538},
year = {2012}
}

@article{GarciaRojas:15,
adsurl = {https://ui.adsabs.harvard.edu/abs/2015MNRAS.452.2606G},
archiveprefix = {arXiv},
author = {{Garc{\'\i}a-Rojas}, J. and {Madonna}, S. and {Luridiana}, V. and {Sterling}, N.~C. and {Morisset}, C. and {Delgado-Inglada}, G. and {Toribio San Cipriano}, L.},
doi = {10.1093/mnras/stv1415},
eprint = {1506.07079},
journal = {\mnras},
month = {September},
number = {3},
pages = {2606-2640},
primaryclass = {astro-ph.SR},
title = {{s-process enrichment in the planetary nebula NGC 3918. Results from deep echelle spectrophotometry}},
volume = {452},
year = {2015}
}

@article{GarciaRojas:18,
adsurl = {https://ui.adsabs.harvard.edu/abs/2018MNRAS.473.4476G},
archiveprefix = {arXiv},
author = {{Garc{\'\i}a-Rojas}, J. and {Delgado-Inglada}, G. and {Garc{\'\i}a-Hern{\'a}ndez}, D.~A. and {Dell'Agli}, F. and {Lugaro}, M. and {Karakas}, A.~I. and {Rodr{\'\i}guez}, M.},
doi = {10.1093/mnras/stx2519},
eprint = {1709.07958},
journal = {\mnras},
month = {February},
number = {4},
pages = {4476-4496},
primaryclass = {astro-ph.SR},
title = {{C/O ratios in planetary nebulae with dual-dust chemistry from faint optical recombination lines}},
volume = {473},
year = {2018}
}

@article{MendezDelgado:21a,
adsurl = {https://ui.adsabs.harvard.edu/abs/2021MNRAS.502.1703M},
archiveprefix = {arXiv},
author = {{M{\'e}ndez-Delgado}, J.~E. and {Esteban}, C. and {Garc{\'\i}a-Rojas}, J. and {Henney}, W.~J. and {Mesa-Delgado}, A. and {Arellano-C{\'o}rdova}, K.~Z.},
doi = {10.1093/mnras/stab068},
eprint = {2101.02191},
journal = {\mnras},
month = {April},
number = {2},
pages = {1703-1739},
primaryclass = {astro-ph.GA},
title = {{Photoionized Herbig-Haro objects in the Orion Nebula through deep high-spectral resolution spectroscopy - I. HH 529 II and III}},
volume = {502},
year = {2021}
}

@article{MendezDelgado:21b,
adsurl = {https://ui.adsabs.harvard.edu/abs/2021ApJ...918...27M},
archiveprefix = {arXiv},
author = {{M{\'e}ndez-Delgado}, J.~E. and {Henney}, W.~J. and {Esteban}, C. and {Garc{\'\i}a-Rojas}, J. and {Mesa-Delgado}, A. and {Arellano-C{\'o}rdova}, K.~Z.},
doi = {10.3847/1538-4357/ac0cf5},
eid = {27},
eprint = {2106.08667},
journal = {\apj},
month = {September},
number = {1},
pages = {27},
primaryclass = {astro-ph.GA},
title = {{Photoionized Herbig-Haro Objects in the Orion Nebula through Deep High Spectral Resolution Spectroscopy. II. HH 204}},
volume = {918},
year = {2021}
}

@article{Sharpee:07,
adsurl = {https://ui.adsabs.harvard.edu/abs/2007ApJ...659.1265S},
archiveprefix = {arXiv},
author = {{Sharpee}, Brian and {Zhang}, Yong and {Williams}, Robert and {Pellegrini}, Eric and {Cavagnolo}, Kenneth and {Baldwin}, Jack A. and {Phillips}, Mark and {Liu}, Xiao-Wei},
doi = {10.1086/511665},
eprint = {astro-ph/0612101},
journal = {\apj},
month = {April},
number = {2},
pages = {1265-1290},
primaryclass = {astro-ph},
title = {{s-Process Abundances in Planetary Nebulae}},
volume = {659},
year = {2007}
}

@article{Sowicka:17,
adsurl = {https://ui.adsabs.harvard.edu/abs/2017MNRAS.471.3529S},
archiveprefix = {arXiv},
author = {{Sowicka}, Paulina and {Jones}, David and {Corradi}, Romano L.~M. and {Wesson}, Roger and {Garc{\'\i}a-Rojas}, Jorge and {Santander-Garc{\'\i}a}, Miguel and {Boffin}, Henri M.~J. and {Rodr{\'\i}guez-Gil}, Pablo},
doi = {10.1093/mnras/stx1697},
eprint = {1706.08766},
journal = {\mnras},
month = {November},
number = {3},
pages = {3529-3546},
primaryclass = {astro-ph.SR},
title = {{The planetary nebula IC 4776 and its post-common-envelope binary central star}},
volume = {471},
year = {2017}
}

@ARTICLE{MendezDelgado:24,
       author = {{M{\'e}ndez-Delgado}, J.~E. and {Kreckel}, K. and {Esteban}, C. and {Garc{\'\i}a-Rojas}, J. and {Carigi}, L. and {Sander}, A.~A.~C. and {Palla}, M. and {Chru{\'s}li{\'n}ska}, M. and {De Looze}, I. and {Rela{\~n}o}, M. and {van der Giessen}, S.~A. and {Reyes-Rodr{\'\i}guez}, E. and {S{\'a}nchez}, S.~F.},
        title = "{Gas-phase Fe/O and Fe/N abundances in Star-Forming Regions. Relations between nucleosynthesis, metallicity and dust}",
      journal = {arXiv e-prints},
         year = 2024,
        month = aug,
          eid = {arXiv:2408.06215},
        pages = {arXiv:2408.06215},
          doi = {10.48550/arXiv.2408.06215},
archivePrefix = {arXiv},
       eprint = {2408.06215},
 primaryClass = {astro-ph.GA},
       adsurl = {https://ui.adsabs.harvard.edu/abs/2024arXiv240806215M}
}

@ARTICLE{Keenan:97,
       author = {{Keenan}, F.~P. and {McKenna}, F.~C. and {Bell}, K.~L. and {Ramsbottom}, C.~A. and {Wickstead}, A.~W. and {Aller}, L.~H. and {Hyung}, S.},
        title = "{Nebular and Auroral Emission Lines of [Ar IV] in the Optical Spectra of Planetary Nebulae}",
      journal = {\apj},
         year = 1997,
        month = sep,
       volume = {487},
       number = {1},
        pages = {457-462},
          doi = {10.1086/304594},
       adsurl = {https://ui.adsabs.harvard.edu/abs/1997ApJ...487..457K}
}

@ARTICLE{Czyzak:80,
       author = {{Czyzak}, S.~J. and {Sonneborn}, G. and {Aller}, L.~H. and {Shectman}, S.~A.},
        title = "{Nebular and auroral transitions of (Ar IV) in some planetary nebulae.}",
      journal = {\apj},
         year = 1980,
        month = oct,
       volume = {241},
        pages = {719-724},
          doi = {10.1086/158382},
       adsurl = {https://ui.adsabs.harvard.edu/abs/1980ApJ...241..719C}
}

@INCOLLECTION{GarciaRojas:20,
       author = {{Garc{\'\i}a-Rojas}, Jorge},
        title = "{Physical Conditions and Chemical Abundances in Photoionized Nebulae from Optical Spectra}",
    booktitle = {Reviews in Frontiers of Modern Astrophysics; From Space Debris to Cosmology},
         year = 2020,
       editor = {{Kab{\'a}th}, Petr and {Jones}, David and {Skarka}, Marek},
        pages = {89-121},
          doi = {10.1007/978-3-030-38509-5_4},
       adsurl = {https://ui.adsabs.harvard.edu/abs/2020rfma.book...89G}
}

@ARTICLE{GarciaRojas:22,
       author = {{Garc{\'\i}a-Rojas}, J. and {Morisset}, C. and {Jones}, D. and {Wesson}, R. and {Boffin}, H.~M.~J. and {Monteiro}, H. and {Corradi}, R.~L.~M. and {Rodr{\'\i}guez-Gil}, P.},
        title = "{MUSE spectroscopy of planetary nebulae with high abundance discrepancies}",
      journal = {\mnras},
         year = 2022,
        month = mar,
       volume = {510},
       number = {4},
        pages = {5444-5463},
          doi = {10.1093/mnras/stab3523},
archivePrefix = {arXiv},
       eprint = {2112.00480},
 primaryClass = {astro-ph.SR},
       adsurl = {https://ui.adsabs.harvard.edu/abs/2022MNRAS.510.5444G}
}

@ARTICLE{GomezLlanos:20b,
       author = {{G{\'o}mez-Llanos}, V. and {Morisset}, C. and {Garc{\'\i}a-Rojas}, J. and {Jones}, D. and {Wesson}, R. and {Corradi}, R.~L.~M. and {Boffin}, H.~M.~J.},
        title = "{The impact of strong recombination on temperature determination in planetary nebulae}",
      journal = {\mnras},
         year = 2020,
        month = nov,
       volume = {498},
       number = {1},
        pages = {L82-L86},
          doi = {10.1093/mnrasl/slaa131},
archivePrefix = {arXiv},
       eprint = {2007.05488},
 primaryClass = {astro-ph.GA},
       adsurl = {https://ui.adsabs.harvard.edu/abs/2020MNRAS.498L..82G}
}

@ARTICLE{GomezLlanos:24,
       author = {{G{\'o}mez-Llanos}, V. and {Garc{\'\i}a-Rojas}, J. and {Morisset}, C. and {Monteiro}, H. and {Jones}, D. and {Wesson}, R. and {Boffin}, H.~M.~J. and {Corradi}, R.~L.~M.},
        title = "{MUSE spectroscopy of the high abundance discrepancy planetary nebula NGC 6153}",
      journal = {\aap},
         year = 2024,
        month = sep,
       volume = {689},
          eid = {A228},
        pages = {A228},
          doi = {10.1051/0004-6361/202450822},
archivePrefix = {arXiv},
       eprint = {2407.06385},
 primaryClass = {astro-ph.SR},
       adsurl = {https://ui.adsabs.harvard.edu/abs/2024A&A...689A.228G}
}

@ARTICLE{Liu:00,
       author = {{Liu}, X. -W. and {Storey}, P.~J. and {Barlow}, M.~J. and {Danziger}, I.~J. and {Cohen}, M. and {Bryce}, M.},
        title = "{NGC 6153: a super-metal-rich planetary nebula?}",
      journal = {\mnras},
         year = 2000,
        month = mar,
       volume = {312},
       number = {3},
        pages = {585-628},
          doi = {10.1046/j.1365-8711.2000.03167.x},
       adsurl = {https://ui.adsabs.harvard.edu/abs/2000MNRAS.312..585L}
}

@ARTICLE{Morisset:15,
       author = {{Morisset}, C. and {Delgado-Inglada}, G. and {Flores-Fajardo}, N.},
        title = "{A virtual observatory for photoionized nebulae: the Mexican Million Models database (3MdB).}",
      journal = {\rmxaa},
         year = 2015,
        month = apr,
       volume = {51},
        pages = {103-120},
          doi = {10.48550/arXiv.1412.5349},
archivePrefix = {arXiv},
       eprint = {1412.5349},
 primaryClass = {astro-ph.GA},
       adsurl = {https://ui.adsabs.harvard.edu/abs/2015RMxAA..51..103M}
}

@ARTICLE{DelgadoInglada:14,
       author = {{Delgado-Inglada}, Gloria and {Morisset}, Christophe and {Stasi{\'n}ska}, Gra{\.z}yna},
        title = "{Ionization correction factors for planetary nebulae - I. Using optical spectra}",
      journal = {\mnras},
         year = 2014,
        month = may,
       volume = {440},
       number = {1},
        pages = {536-554},
          doi = {10.1093/mnras/stu341},
archivePrefix = {arXiv},
       eprint = {1402.4852},
 primaryClass = {astro-ph.SR},
       adsurl = {https://ui.adsabs.harvard.edu/abs/2014MNRAS.440..536D}
}

@ARTICLE{Sabbadin:77,
       author = {{Sabbadin}, F. and {Minello}, S. and {Bianchini}, A.},
        title = "{Sharpless 176: a large, nearby planetary nebula.}",
      journal = {\aap},
         year = 1977,
        month = aug,
       volume = {60},
        pages = {147-149},
       adsurl = {https://ui.adsabs.harvard.edu/abs/1977A&A....60..147S}
}

@ARTICLE{Sabin:13,
       author = {{Sabin}, L. and {Parker}, Q.~A. and {Contreras}, M.~E. and {Olgu{\'\i}n}, L. and {Frew}, D.~J. and {Stupar}, M. and {V{\'a}zquez}, R. and {Wright}, N.~J. and {Corradi}, R.~L.~M. and {Morris}, R.~A.~H.},
        title = "{New Galactic supernova remnants discovered with IPHAS}",
      journal = {\mnras},
         year = 2013,
        month = may,
       volume = {431},
       number = {1},
        pages = {279-291},
          doi = {10.1093/mnras/stt160},
archivePrefix = {arXiv},
       eprint = {1301.6416},
 primaryClass = {astro-ph.SR},
       adsurl = {https://ui.adsabs.harvard.edu/abs/2013MNRAS.431..279S}
}

@INPROCEEDINGS{Garciarojas:25,
       author = {{Garc{\'\i}a-Rojas}, J. and {Reyes-Rodr{\'\i}guez}, E. and {M{\'e}ndez-Delgado}, J.~E. and {Esteban}, C. and {Orte-Garc{\'\i}a}, M.},
        title = "{DESIRED: A Database of High-Quality Spectra to Study Fundamental Problems in Ionized Nebulae}",
    booktitle = {Highlights of Spanish Astrophysics XII},
         year = 2025,
       editor = {{Manteiga}, M. and {Gonz{\'a}lez-Galindo}, F. and {Labiano-Ortega}, A. and {Mart{\'\i}nez-Gonz{\'a}lez}, M.~J. and {Rea}, N. and {Romero-G{\'o}mez}, M. and {Ulla-Miguel}, A. and {Yepes}, G. and {Rodr{\'\i}guez-L{\'o}pez}, C. and {G{\'o}mez-Garc{\'\i}a}, A. and {Dafonte}, C.},
        month = may,
        pages = {216},
       adsurl = {https://ui.adsabs.harvard.edu/abs/2025hsa..conf..216G}
}

@ARTICLE{JuandeDios:17,
       author = {{Juan de Dios}, Leticia and {Rodr{\'\i}guez}, M{\'o}nica},
        title = "{The impact of atomic data selection on nebular abundance determinations}",
      journal = {\mnras},
         year = 2017,
        month = jul,
       volume = {469},
       number = {1},
        pages = {1036-1053},
          doi = {10.1093/mnras/stx916},
archivePrefix = {arXiv},
       eprint = {1704.06009},
 primaryClass = {astro-ph.GA},
       adsurl = {https://ui.adsabs.harvard.edu/abs/2017MNRAS.469.1036J}
}

@ARTICLE{Pottasch:99,
       author = {{Pottasch}, S.~R. and {Beintema}, D.~A.},
        title = "{The ISO spectrum of the planetary nebula NGC 6302. II. Nebular abundances}",
      journal = {\aap},
         year = 1999,
        month = jul,
       volume = {347},
        pages = {975-982},
       adsurl = {https://ui.adsabs.harvard.edu/abs/1999A&A...347..975P}
}

@ARTICLE{Pottasch:00,
       author = {{Pottasch}, S.~R. and {Beintema}, D.~A. and {Feibelman}, W.~A.},
        title = "{Abundance in the planetary nebulae NGC 6537 and He2-111}",
      journal = {\aap},
         year = 2000,
        month = nov,
       volume = {363},
        pages = {767-778},
       adsurl = {https://ui.adsabs.harvard.edu/abs/2000A&A...363..767P}
}

@ARTICLE{Pottasch:07,
       author = {{Pottasch}, S.~R. and {Surendiranath}, R.},
        title = "{Abundances in planetary nebulae: <ASTROBJ>Hb 5</ASTROBJ>}",
      journal = {\aap},
         year = 2007,
        month = jan,
       volume = {462},
       number = {1},
        pages = {179-192},
          doi = {10.1051/0004-6361:20066012},
       adsurl = {https://ui.adsabs.harvard.edu/abs/2007A&A...462..179P}
}

@ARTICLE{Pottasch:08a,
       author = {{Pottasch}, S.~R. and {Bernard-Salas}, J. and {Roellig}, T.~L.},
        title = "{Abundances of planetary nebula NGC 2392}",
      journal = {\aap},
         year = 2008,
        month = apr,
       volume = {481},
       number = {2},
        pages = {393-400},
          doi = {10.1051/0004-6361:20079041},
archivePrefix = {arXiv},
       eprint = {0801.2767},
 primaryClass = {astro-ph},
       adsurl = {https://ui.adsabs.harvard.edu/abs/2008A&A...481..393P}
}

@ARTICLE{Pottasch:08b,
       author = {{Pottasch}, S.~R. and {Bernard-Salas}, J.},
        title = "{Abundances of planetary nebulae NGC 3242 and NGC 6369}",
      journal = {\aap},
         year = 2008,
        month = nov,
       volume = {490},
       number = {2},
        pages = {715-724},
          doi = {10.1051/0004-6361:200810721},
archivePrefix = {arXiv},
       eprint = {0809.3745},
 primaryClass = {astro-ph},
       adsurl = {https://ui.adsabs.harvard.edu/abs/2008A&A...490..715P}
}

@ARTICLE{Pottasch:09,
       author = {{Pottasch}, S.~R. and {Surendiranath}, R. and {Bernard-Salas}, J. and {Roellig}, T.~L.},
        title = "{Abundances in planetary nebulae: NGC 2792}",
      journal = {\aap},
         year = 2009,
        month = jul,
       volume = {502},
       number = {1},
        pages = {189-197},
          doi = {10.1051/0004-6361/200912076},
       adsurl = {https://ui.adsabs.harvard.edu/abs/2009A&A...502..189P}
}

@ARTICLE{Zeippen:87,
       author = {{Zeippen}, C.~J. and {Butler}, K. and {Le Bourlot}, J.},
        title = "{Effective collision strengths for fine-structure forbidden transitions in the 3p3 configuration of AR IV}",
      journal = {\aap},
         year = 1987,
        month = dec,
       volume = {188},
       number = {1},
        pages = {251-257},
       adsurl = {https://ui.adsabs.harvard.edu/abs/1987A&A...188..251Z}
}

@ARTICLE{Czyzak:63,
       author = {{Czyzak}, S.~J. and {Krueger}, T.~K.},
        title = "{Forbidden transition probabilities for some P, S, CI and A ions}",
      journal = {\mnras},
         year = 1963,
        month = jan,
       volume = {126},
        pages = {177},
          doi = {10.1093/mnras/126.2.177},
       adsurl = {https://ui.adsabs.harvard.edu/abs/1963MNRAS.126..177C}
}

@ARTICLE{Rynkun:19,
       author = {{Rynkun}, P. and {Gaigalas}, G. and {J{\"o}nsson}, P.},
        title = "{Theoretical investigation of energy levels and transition data for S II, Cl III, Ar IV}",
      journal = {\aap},
         year = 2019,
        month = mar,
       volume = {623},
          eid = {A155},
        pages = {A155},
          doi = {10.1051/0004-6361/201834931},
       adsurl = {https://ui.adsabs.harvard.edu/abs/2019A&A...623A.155R}
}

@ARTICLE{Mendoza:82a,
       author = {{Mendoza}, C. and {Zeippen}, C.~J.},
        title = "{Transition probabilities for forbidden lines in the 3p3 configuration}",
      journal = {\mnras},
         year = 1982,
        month = jan,
       volume = {198},
        pages = {127-139},
          doi = {10.1093/mnras/198.1.127},
       adsurl = {https://ui.adsabs.harvard.edu/abs/1982MNRAS.198..127M}
}

@ARTICLE{Morisset:20,
       author = {{Morisset}, Christophe and {Luridiana}, Valentina and {Garc{\'\i}a-Rojas}, Jorge and {G{\'o}mez-Llanos}, Ver{\'o}nica and {Bautista}, Manuel and {Mendoza}, Claudio},
        title = "{Atomic Data Assessment with PyNeb}",
      journal = {Atoms},
         year = 2020,
        month = oct,
       volume = {8},
       number = {4},
          eid = {66},
        pages = {66},
          doi = {10.3390/atoms8040066},
archivePrefix = {arXiv},
       eprint = {2009.10586},
 primaryClass = {astro-ph.GA},
       adsurl = {https://ui.adsabs.harvard.edu/abs/2020Atoms...8...66M}
}

@ARTICLE{Wang:07,
       author = {{Wang}, W. and {Liu}, X.-W.},
        title = "{Elemental abundances of Galactic bulge planetary nebulae from optical recombination lines}",
      journal = {\mnras},
         year = 2007,
        month = oct,
       volume = {381},
       number = {2},
        pages = {669-701},
          doi = {10.1111/j.1365-2966.2007.12198.x},
archivePrefix = {arXiv},
       eprint = {0707.0542},
 primaryClass = {astro-ph},
       adsurl = {https://ui.adsabs.harvard.edu/abs/2007MNRAS.381..669W}
}

@ARTICLE{Liu:04,
       author = {{Liu}, Y. and {Liu}, X.-W. and {Luo}, S.-G. and {Barlow}, M.~J.},
        title = "{Chemical abundances of planetary nebulae from optical recombination lines ─ I. Observations and plasma diagnostics}",
      journal = {\mnras},
         year = 2004,
        month = oct,
       volume = {353},
       number = {4},
        pages = {1231-1250},
          doi = {10.1111/j.1365-2966.2004.08155.x},
       adsurl = {https://ui.adsabs.harvard.edu/abs/2004MNRAS.353.1231L}
}

@ARTICLE{Hyung:01a,
       author = {{Hyung}, Siek and {Aller}, Lawrence H. and {Lee}, Woo-baik},
        title = "{Spectroscopic Observation of the Planetary Nebula IC 4846}",
      journal = {\pasp},
         year = 2001,
        month = dec,
       volume = {113},
       number = {790},
        pages = {1559-1568},
          doi = {10.1086/324415},
       adsurl = {https://ui.adsabs.harvard.edu/abs/2001PASP..113.1559H}
}

@ARTICLE{Tan:24,
       author = {{Tan}, Shuyu and {Parker}, Quentin A. and {Zijlstra}, Albert A. and {Rees}, Bryan},
        title = "{A catalogue of planetary nebulae chemical abundances in the Galactic bulge}",
      journal = {\mnras},
         year = 2024,
        month = jan,
       volume = {527},
       number = {3},
        pages = {6363-6387},
          doi = {10.1093/mnras/stad3496},
archivePrefix = {arXiv},
       eprint = {2311.01836},
 primaryClass = {astro-ph.GA},
       adsurl = {https://ui.adsabs.harvard.edu/abs/2024MNRAS.527.6363T}
}

@ARTICLE{Ercolano:04,
       author = {{Ercolano}, B. and {Wesson}, R. and {Zhang}, Y. and {Barlow}, M.~J. and {De Marco}, O. and {Rauch}, T. and {Liu}, X.-W.},
        title = "{Observations and three-dimensional photoionization modelling of the Wolf-Rayet planetary nebula NGC 1501}",
      journal = {\mnras},
         year = 2004,
        month = oct,
       volume = {354},
       number = {2},
        pages = {558-574},
          doi = {10.1111/j.1365-2966.2004.08218.x},
archivePrefix = {arXiv},
       eprint = {astro-ph/0407230},
 primaryClass = {astro-ph},
       adsurl = {https://ui.adsabs.harvard.edu/abs/2004MNRAS.354..558E}
}

@ARTICLE{Ruiz:03,
       author = {{Ruiz}, Mar{\'\i}a Teresa and {Peimbert}, Antonio and {Peimbert}, Manuel and {Esteban}, C{\'e}sar},
        title = "{Very Large Telescope Echelle Spectrophotometry of the Planetary Nebula NGC 5307 and Temperature Variations}",
      journal = {\apj},
         year = 2003,
        month = sep,
       volume = {595},
       number = {1},
        pages = {247-258},
          doi = {10.1086/377255},
archivePrefix = {arXiv},
       eprint = {astro-ph/0305348},
 primaryClass = {astro-ph},
       adsurl = {https://ui.adsabs.harvard.edu/abs/2003ApJ...595..247R}
}

@ARTICLE{Hyung:94b,
       author = {{Hyung}, S. and {Aller}, L.~H. and {Feibelman}, W.~A.},
        title = "{The spectrum of the planetary nebula NGC 6572.}",
      journal = {\mnras},
         year = 1994,
        month = aug,
       volume = {269},
        pages = {975-997},
          doi = {10.1093/mnras/269.4.975},
       adsurl = {https://ui.adsabs.harvard.edu/abs/1994MNRAS.269..975H}
}

@ARTICLE{Hyung:97a,
       author = {{Hyung}, Siek and {Aller}, Lawrence H.},
        title = "{The high-excitation planetary nebula NGC 6741}",
      journal = {\mnras},
         year = 1997,
        month = nov,
       volume = {292},
       number = {1},
        pages = {71-85},
          doi = {10.1093/mnras/292.1.71},
       adsurl = {https://ui.adsabs.harvard.edu/abs/1997MNRAS.292...71H}
}

@ARTICLE{Hyung:97b,
       author = {{Hyung}, Siek and {Aller}, Lawrence H.},
        title = "{The High-Excitation Planetary Nebula NGC 7662}",
      journal = {\apj},
         year = 1997,
        month = dec,
       volume = {491},
       number = {1},
        pages = {242-253},
          doi = {10.1086/304948},
       adsurl = {https://ui.adsabs.harvard.edu/abs/1997ApJ...491..242H}
}

@ARTICLE{GarciaRojas:09,
       author = {{Garc{\'\i}a-Rojas}, J. and {Pe{\~n}a}, M. and {Peimbert}, A.},
        title = "{Faint recombination lines in Galactic PNe with a [WC] nucleus}",
      journal = {\aap},
         year = 2009,
        month = mar,
       volume = {496},
       number = {1},
        pages = {139-152},
          doi = {10.1051/0004-6361:200811185},
archivePrefix = {arXiv},
       eprint = {0812.3049},
 primaryClass = {astro-ph},
       adsurl = {https://ui.adsabs.harvard.edu/abs/2009A&A...496..139G}
}

@ARTICLE{Fang:11,
       author = {{Fang}, X. and {Liu}, X.-W.},
        title = "{Very deep spectroscopy of the bright Saturn nebula NGC 7009 - I. Observations and plasma diagnostics}",
      journal = {\mnras},
         year = 2011,
        month = jul,
       volume = {415},
       number = {1},
        pages = {181-198},
          doi = {10.1111/j.1365-2966.2011.18681.x},
archivePrefix = {arXiv},
       eprint = {1103.1705},
 primaryClass = {astro-ph.SR},
       adsurl = {https://ui.adsabs.harvard.edu/abs/2011MNRAS.415..181F}
}

@ARTICLE{numpy,  
       author = {{Van Der Walt}, St{\'e}fan and {Colbert}, S. Chris and {Varoquaux}, Ga{\"e}l},
        title = "{The NumPy Array: A Structure for Efficient Numerical Computation}",
      journal = {Computing in Science and Engineering},
         year = 2011,
        month = mar,
       volume = {13},
       number = {2},
        pages = {22-30},
          doi = {10.1109/MCSE.2011.37},
archivePrefix = {arXiv},
       eprint = {1102.1523},
 primaryClass = {cs.MS},
       adsurl = {https://ui.adsabs.harvard.edu/abs/2011CSE....13b..22V}
}

@ARTICLE{scipy, 
       author = {{Virtanen}, Pauli and {Gommers}, Ralf and {Oliphant}, Travis E. and {Haberland}, Matt and {Reddy}, Tyler and {Cournapeau}, David and {Burovski}, Evgeni and {Peterson}, Pearu and {Weckesser}, Warren and {Bright}, Jonathan and {van der Walt}, St{\'e}fan J. and {Brett}, Matthew and {Wilson}, Joshua and {Millman}, K. Jarrod and {Mayorov}, Nikolay and {Nelson}, Andrew R.~J. and {Jones}, Eric and {Kern}, Robert and {Larson}, Eric and {Carey}, C.~J. and {Polat}, {\.I}lhan and {Feng}, Yu and {Moore}, Eric W. and {VanderPlas}, Jake and {Laxalde}, Denis and {Perktold}, Josef and {Cimrman}, Robert and {Henriksen}, Ian and {Quintero}, E.~A. and {Harris}, Charles R. and {Archibald}, Anne M. and {Ribeiro}, Ant{\^o}nio H. and {Pedregosa}, Fabian and {van Mulbregt}, Paul and {SciPy 1.  0 Contributors}},
        title = "{SciPy 1.0: fundamental algorithms for scientific computing in Python}",
      journal = {Nature Medicine},
         year = 2020,
        month = feb,
       volume = {17},
        pages = {261-272},
          doi = {10.1038/s41592-019-0686-2},
archivePrefix = {arXiv},
       eprint = {1907.10121},
 primaryClass = {cs.MS},
       adsurl = {https://ui.adsabs.harvard.edu/abs/2020NatMe..17..261V}
}

@ARTICLE{matplotlib,  
       author = {{Hunter}, John D.},
        title = "{Matplotlib: A 2D Graphics Environment}",
      journal = {Computing in Science and Engineering},
         year = 2007,
        month = jan,
       volume = {9},
       number = {3},
        pages = {90-95},
          doi = {10.1109/MCSE.2007.55},
       adsurl = {https://ui.adsabs.harvard.edu/abs/2007CSE.....9...90H}
}

@incollection{jupyter, 
       author = {{Kluyver}, Thomas and {Ragan-Kelley}, Benjain and {P{\'e}rez}, Fernando and {Granger}, Brian and {Bussonnier}, Matthias and {Frederic}, Jonathan and {Kelley}, Kyle and {Hamrick}, Jessica and {Grout}, Jason and {Corlay}, Sylvain and {Ivanov}, Paul and {Avila}, Dami{\'a}n and {Abdalla}, Safia and {Willing}, Carol and {Jupyter Development Team}},
        title = "{Jupyter Notebooks{\textemdash}a publishing format for reproducible computational workflows}",
    booktitle = {IOS Press},
         year = 2016,
        pages = {87-90},
          doi = {10.3233/978-1-61499-649-1-87},
       adsurl = {https://ui.adsabs.harvard.edu/abs/2016ppap.book...87K}
}

@InProceedings{pandas:10,
  author    = {{McKinney}, Wes},
  title     = { {D}ata {S}tructures for {S}tatistical {C}omputing in {P}ython },
  booktitle = { {P}roceedings of the 9th {P}ython in {S}cience {C}onference },
  pages     = { 56 - 61 },
  year      = { 2010 },
  editor    = { {S}t\'efan van der {W}alt and {J}arrod {M}illman },
  doi       = { 10.25080/Majora-92bf1922-00a }
}

@InProceedings{pandas:20,
  author    = {{McKinney}, Wes},
  title     = {Data Structures for Statistical Computing in Python},
  booktitle = {Proceedings of the 9th Python in Science Conference},
  pages     = {51-56},
  year      = {2020},
  editor    = {St\'efan van der Walt and Jarrod Millman}
}

@ARTICLE{Dopita:00,
       author = {{Dopita}, Michael A. and {Sutherland}, Ralph S.},
        title = "{The Importance of Photoelectric Heating by Dust in Planetary Nebulae}",
      journal = {\apj},
         year = 2000,
        month = aug,
       volume = {539},
       number = {2},
        pages = {742-750},
          doi = {10.1086/309241},
       adsurl = {https://ui.adsabs.harvard.edu/abs/2000ApJ...539..742D}
}

@ARTICLE{Stasinska:02,
       author = {{Stasi{\'n}ska}, Grazyna},
        title = "{Abundance determinations in HII regions and planetary nebulae}",
      journal = {arXiv e-prints},
         year = 2002,
        month = jul,
          eid = {astro-ph/0207500},
        pages = {astro-ph/0207500},
          doi = {10.48550/arXiv.astro-ph/0207500},
archivePrefix = {arXiv},
       eprint = {astro-ph/0207500},
 primaryClass = {astro-ph},
       adsurl = {https://ui.adsabs.harvard.edu/abs/2002astro.ph..7500S}
}

@ARTICLE{Kwitter:03,
       author = {{Kwitter}, K.~B. and {Henry}, R.~B.~C. and {Milingo}, J.~B.},
        title = "{Sulfur, Chlorine, and Argon Abundances in Planetary Nebulae. III. Observations and Results for a Final Sample}",
      journal = {\pasp},
         year = 2003,
        month = jan,
       volume = {115},
       number = {803},
        pages = {80-95},
          doi = {10.1086/345108},
archivePrefix = {arXiv},
       eprint = {astro-ph/0209543},
 primaryClass = {astro-ph},
       adsurl = {https://ui.adsabs.harvard.edu/abs/2003PASP..115...80K}
}

@ARTICLE{Henry:10,
       author = {{Henry}, R.~B.~C. and {Kwitter}, Karen B. and {Jaskot}, Anne E. and {Balick}, Bruce and {Morrison}, Michael A. and {Milingo}, Jacquelynne B.},
        title = "{Abundances of Galactic Anticenter Planetary Nebulae and the Oxygen Abundance Gradient in the Galactic Disk}",
      journal = {\apj},
         year = 2010,
        month = nov,
       volume = {724},
       number = {1},
        pages = {748-761},
          doi = {10.1088/0004-637X/724/1/748},
archivePrefix = {arXiv},
       eprint = {1009.1921},
 primaryClass = {astro-ph.GA},
       adsurl = {https://ui.adsabs.harvard.edu/abs/2010ApJ...724..748H}
}

@ARTICLE{Milingo:02,
       author = {{Milingo}, J.~B. and {Kwitter}, K.~B. and {Henry}, R.~B.~C. and {Cohen}, R.~E.},
        title = "{Sulfur, Chlorine, and Argon in Planetary Nebulae. IIA. Observations of a Southern Sample}",
      journal = {\apjs},
         year = 2002,
        month = feb,
       volume = {138},
       number = {2},
        pages = {279-283},
          doi = {10.1086/324291},
archivePrefix = {arXiv},
       eprint = {astro-ph/0108336},
 primaryClass = {astro-ph},
       adsurl = {https://ui.adsabs.harvard.edu/abs/2002ApJS..138..279M}
}

@ARTICLE{Milingo:10,
       author = {{Milingo}, J.~B. and {Kwitter}, K.~B. and {Henry}, R.~B.~C. and {Souza}, S.~P.},
        title = "{Alpha Element Abundances in a Large Sample of Galactic Planetary Nebulae}",
      journal = {\apj},
         year = 2010,
        month = mar,
       volume = {711},
       number = {2},
        pages = {619-630},
          doi = {10.1088/0004-637X/711/2/619},
archivePrefix = {arXiv},
       eprint = {1002.3606},
 primaryClass = {astro-ph.SR},
       adsurl = {https://ui.adsabs.harvard.edu/abs/2010ApJ...711..619M}
}

@ARTICLE{Hyung:97c,
       author = {{Hyung}, Siek and {Aller}, Lawrence H. and {Feibelman}, Walter A.},
        title = "{The Spectrum of the Planetary Nebula NGC 6884}",
      journal = {\apjs},
         year = 1997,
        month = feb,
       volume = {108},
       number = {2},
        pages = {503-513},
          doi = {10.1086/312969},
       adsurl = {https://ui.adsabs.harvard.edu/abs/1997ApJS..108..503H}
}

@ARTICLE{Feibelman:96,
       author = {{Feibelman}, Walter A. and {Hyung}, Siek and {Aller}, Lawrence H.},
        title = "{The spectrum of the planetary nebula IC 351}",
      journal = {\mnras},
         year = 1996,
        month = jan,
       volume = {278},
       number = {2},
        pages = {625-634},
          doi = {10.1093/mnras/278.2.625},
       adsurl = {https://ui.adsabs.harvard.edu/abs/1996MNRAS.278..625F}
}

\clearpage

\begin{appendix} 
\onecolumn

\section{Table of references and physical conditions}
\label{sec:appendix_a}

\nopagebreak 

\footnotesize
\renewcommand{\arraystretch}{1.2}
\begin{longtable}{cccccccc}
\caption{PN names, physical conditions computed from DESIRED-E and original data references for our PN sample.}\\
\hline
\hline
Galaxy & PN & \nel(average)& \nel({\ariv}) & \nel(\feiii) & \tel(\oiii) & \tel(\ariv)  & Ref.  \\
& Name$^{\rm a}$ & [cm$^{-3}$] & [cm$^{-3}$] & [cm$^{-3}$] & [K] & [K] & \\
\hline
\endfirsthead
\caption{Continued.}\\
\hline
Galaxy & PN & \nel(average)& \nel({\ariv}) & \nel(\feiii) & \tel(\oiii) & \tel(\ariv)  & Ref.  \\
& Name$^{\rm a}$ & [cm$^{-3}$] & [cm$^{-3}$] & [cm$^{-3}$] & [K] & [K] & \\
\hline
\endhead
\hline
\endfoot
\hline
\endlastfoot
\label{table:physical_conditions}
LMC & SMP~73 & $10610 \pm 1180$ & $10840 ^{+1000} _{-810}$ &  -  & $12270 ^{+160} _{-190}$ & $19970 ^{+1500} _{-1700}$ &Mampaso et al. in prep.\\
LMC & SMP~92 & $8790 \pm 1970$ & $9470 ^{+770} _{-990}$ &  -  & $13040 ^{+200} _{-240}$ & $15540 ^{+740} _{-1130}$ &Mampaso et al. in prep.\\
MilkyWay & Cn~2-1 & $9440 \pm 4960$ & $20000 ^{+2200} _{-2840}$ &  -  & $10160 ^{+110} _{-190}$ & $10600 ^{+480} _{-410}$ &\citet{Wang:07}\\
MilkyWay & Hb~4 & $7170 \pm 520$ & $7380 ^{+1250} _{-1400}$ &  -  & $9920 ^{+210} _{-270}$ & $12960 ^{+930} _{-1090}$ &\citet{GarciaRojas:12}\\
MilkyWay & He~2-86 & $25090 \pm 9650$ & $36830 ^{+5630} _{-6050}$ & $39050 ^{+32340} _{-15570}$ & $8350 ^{+170} _{-130}$ & $10160 ^{+850} _{-830}$ &\citet{GarciaRojas:12}\\
MilkyWay & IC351 & $750 ^{+750} _{-400}$ & $1520 ^{+840} _{-650}$ &  -  & $12880 ^{+460} _{-320}$ & $18390 ^{+2490} _{-1410}$ &\citet{Feibelman:96}\\
MilkyWay & IC~2501 & $9360 \pm 970$ & $8730 ^{+680} _{-870}$ &  -  & $9450 ^{+90} _{-120}$ & $9220 ^{+580} _{-690}$ &\citet{Sharpee:07}\\
MilkyWay & IC~3568 & $620 \pm 730$ & $1500 ^{+670} _{-710}$ &  -  & $11210 ^{+170} _{-200}$ & $13230 ^{+1040} _{-910}$ &\citet{Liu:04}\\
MilkyWay & IC~4191 & $11790 \pm 2090$ & $11490 ^{+960} _{-1000}$ &  -  & $9710 ^{+110} _{-90}$ & $11330 \pm 310$ &\citet{Sharpee:07}\\
MilkyWay & IC~4776 & $28420 \pm 11090$ & $37560 ^{+3120} _{-3760}$ & $19080 ^{+25190} _{-8570}$ & $9930 ^{+210} _{-170}$ & $11940 ^{+1070} _{-1690}$ &\citet{Sowicka:17}\\
MilkyWay & IC~4846 & $6720 \pm 2730$ & $7830 ^{+4790} _{-4590}$ &  -  & $10450 ^{+320} _{-390}$ & $16300 ^{+5660} _{-4160}$ &\citet{Hyung:01a}\\
MilkyWay & M~1-60 & $11220 \pm 5590$ & $18210 ^{+3590} _{-3470}$ &  -  & $8760 ^{+150} _{-130}$ & $8370 ^{+800} _{-860}$ &\citet{GarciaRojas:18}\\
MilkyWay & M~1-61 & $21060 \pm 7240$ & $34530 ^{+9280} _{-8090}$ & $14480 ^{+54850} _{-11380}$ & $9120 \pm 210$ & $10930 ^{+1460} _{-1110}$ &\citet{GarciaRojas:12}\\
MilkyWay & M~2-31 & $6610 \pm 1780$ & $7900 ^{+1820} _{-1810}$ &  -  & $9950 ^{+190} _{-250}$ & $10490 ^{+890} _{-960}$ &\citet{GarciaRojas:18}\\
MilkyWay & M~2-36 & $3120 \pm 720$ & $4120 ^{+820} _{-1000}$ &  -  & $8380 ^{+100} _{-80}$ & $10990 ^{+1060} _{-820}$ &\citet{Espiritu:21}\\
MilkyWay & M~3-21 & $13870 \pm 5090$ & $23640 ^{+2800} _{-3230}$ & $33800 ^{+39310} _{-16420}$ & $9840 ^{+160} _{-120}$ & $13450 ^{+580} _{-430}$ &\citet{Wang:07}\\
MilkyWay & NGC~1501 & $940 ^{+260} _{-220}$ & $510 ^{+470} _{-290}$ &  -  & $11070 ^{+170} _{-220}$ & $18080 ^{+4250} _{-2480}$ &\citet{Ercolano:04}\\
MilkyWay & NGC~2022 & $840 ^{+270} _{-230}$ & $1520 ^{+620} _{-700}$ &  -  & $14890 ^{+340} _{-320}$ & $18430 ^{+940} _{-680}$ &\citet{Tsamis:03}\\
MilkyWay & NGC~2440 & $4130 \pm 1210$ & $5190 ^{+620} _{-680}$ &  -  & $14530 ^{+290} _{-200}$ & $22570 ^{+900} _{-620}$ &\citet{Sharpee:07}\\
MilkyWay & NGC~3242 & $1690 \pm 320$ & $2180 ^{+600} _{-730}$ &  -  & $11730 ^{+180} _{-300}$ & $12570 ^{+590} _{-450}$ &\citet{Tsamis:03}\\
MilkyWay & NGC~3918 & $6500 \pm 580$ & $6180 ^{+970} _{-1240}$ &  -  & $12680 ^{+320} _{-290}$ & $18010 ^{+1010} _{-1090}$ &\citet{GarciaRojas:15}\\
MilkyWay & NGC~5189 & $1250 \pm 130$ & $1320 ^{+630} _{-610}$ &  -  & $11530 ^{+280} _{-270}$ & $15650 ^{+920} _{-930}$ &\citet{GarciaRojas:12}\\
MilkyWay & NGC~5307 & $2580 \pm 1340$ & $1800 ^{+650} _{-590}$ &  -  & $11680 ^{+90} _{-80}$ & $16720 ^{+2050} _{-1920}$ &\citet{Ruiz:03}\\
MilkyWay & NGC~5882 & $4170 \pm 620$ & $4490 ^{+730} _{-790}$ &  -  & $9330 ^{+130} _{-150}$ & $14470 ^{+1070} _{-1030}$ &\citet{Tsamis:03}\\
MilkyWay & NGC~6210 & $4120 \pm 1250$ & $6400 ^{+980} _{-1160}$ & $14240 ^{+31680} _{-11200}$ & $9490 ^{+200} _{-140}$ & $17750 ^{+1760} _{-1490}$ &\citet{Liu:04}\\
MilkyWay & NGC~6302 & $14800 \pm 5780$ & $14530 ^{+2080} _{-2510}$ &  -  & $18390 ^{+420} _{-540}$ & $24820 ^{+2200} _{-2650}$ &\citet{Tsamis:03}\\
MilkyWay & NGC~6369 & $4130 \pm 540$ & $5020 ^{+1220} _{-1090}$ &  -  & $10620 ^{+180} _{-200}$ & $13690 ^{+1400} _{-1010}$ &\citet{GarciaRojas:12}\\
MilkyWay & NGC~6439 & $5530 \pm 690$ & $5990 ^{+890} _{-1020}$ &  -  & $10260 ^{+130} _{-180}$ & $14870 ^{+420} _{-570}$ &\citet{Wang:07}\\
MilkyWay & NGC~6572 & $24430 \pm 6210$ & $29160 ^{+7580} _{-7490}$ &  -  & $10420 ^{+330} _{-390}$ & $11510 ^{+2250} _{-1670}$ &\citet{Hyung:94b}\\
MilkyWay & NGC~6620 & $2520 \pm 470$ & $2250 ^{+610} _{-580}$ &  -  & $9500 ^{+150} _{-120}$ & $13720 ^{+620} _{-870}$ &\citet{Wang:07}\\
MilkyWay & NGC~6720 & $540 ^{+160} _{-140}$ & $780 ^{+440} _{-490}$ & $9000 ^{+26520} _{-8100}$ & $10440 ^{+200} _{-220}$ & $12930 ^{+1420} _{-1050}$ &\citet{Liu:04}\\
MilkyWay & NGC~6741 & $6380 \pm 1680$ & $5840 ^{+2470} _{-2810}$ &  -  & $12460 ^{+790} _{-600}$ & $19570 ^{+3350} _{-2840}$ &\citet{Hyung:97a}\\
MilkyWay & NGC~6818 & $1890 \pm 450$ & $1850 ^{+770} _{-720}$ &  -  & $13240 ^{+300} _{-330}$ & $15050 ^{+510} _{-520}$ &\citet{Tsamis:03}\\
MilkyWay & NGC6884 & $8150 \pm 2050$ & $6200 ^{+3220} _{-2340}$ &  -  & $9720 ^{+170} _{-120}$ & $14730 ^{+2480} _{-1070}$ &\citet{Hyung:97c}\\
MilkyWay & NGC~7009 & $100 \pm 100$ & $4350 ^{+230} _{-290}$ &  -  & $9880 ^{+160} _{-180}$ & $14250 ^{+760} _{-930}$ &\citet{Fang:11}\\
MilkyWay & NGC~7662 & $3690 \pm 860$ & $3110 ^{+960} _{-1190}$ &  -  & $12440 ^{+580} _{-440}$ & $25900 \pm 2540$ &\citet{Hyung:97b}\\
MilkyWay & PB~6-2 & $2340 \pm 580$ & $1740 ^{+700} _{-720}$ &  -  & $15450 ^{+400} _{-410}$ & $19780 ^{+4110} _{-3400}$ &\citet{GarciaRojas:09}\\
MilkyWay & PC~14 & $3930 \pm 670$ & $4690 ^{+880} _{-980}$ &  -  & $9250 ^{+140} _{-180}$ & $12330 ^{+1740} _{-1870}$ &\citet{GarciaRojas:12}\\
MilkyWay & PNG~000.1+04.3 & $14140 \pm 810$ & $14090 ^{+1280} _{-1570}$ &  -  & $11720 ^{+210} _{-190}$ & $11880 ^{+1190} _{-1060}$ &\citet{Tan:24}\\
MilkyWay & PNG~000.1-02.3 & $1230 \pm 420$ & $1740 ^{+560} _{-390}$ &  -  & $13170 ^{+280} _{-340}$ & $14230 ^{+860} _{-950}$ &\citet{Tan:24}\\
MilkyWay & PNG~000.7-02.7 & $320 ^{+170} _{-120}$ & $6220 ^{+860} _{-750}$ &  -  & $12800 ^{+230} _{-240}$ & $12030 ^{+1300} _{-1440}$ &\citet{Tan:24}\\
MilkyWay & PNG~000.9-04.8 & $1100 \pm 330$ & $1570 ^{+430} _{-470}$ &  -  & $13250 ^{+300} _{-330}$ & $11000 ^{+640} _{-520}$ &\citet{Tan:24}\\
MilkyWay & PNG~001.7+05.7 & $830 ^{+260} _{-160}$ & $1570 ^{+430} _{-380}$ &  -  & $15840 ^{+360} _{-390}$ & $14710 ^{+1060} _{-970}$ &\citet{Tan:24}\\
MilkyWay & PNG~002.9-03.9 & $1510 \pm 470$ & $1030 ^{+470} _{-530}$ &  -  & $13210 ^{+230} _{-360}$ & $20850 ^{+1840} _{-2000}$ &\citet{Tan:24}\\
MilkyWay & PNG~003.8-04.3 & $1520 \pm 70$ & $1430 ^{+270} _{-350}$ &  -  & $10640 ^{+220} _{-190}$ & $13720 ^{+820} _{-680}$ &\citet{Tan:24}\\
MilkyWay & PNG~004.2-03.2 & $920 ^{+770} _{-500}$ & $600 ^{+380} _{-370}$ &  -  & $11320 ^{+290} _{-220}$ & $11960 ^{+1810} _{-2000}$ &\citet{Tan:24}\\
MilkyWay & PNG~005.8-06.1 & $2700 \pm 470$ & $2410 ^{+330} _{-430}$ &  -  & $9420 ^{+100} _{-160}$ & $10200 ^{+590} _{-500}$ &\citet{Tan:24}\\
MilkyWay & PNG~006.4-04.6 & $760 \pm 230$ & $910 \pm 430$ &  -  & $12690 ^{+270} _{-350}$ & $12770 ^{+980} _{-1040}$ &\citet{Tan:24}\\
MilkyWay & PNG~006.8+02.3 & $2660 \pm 1200$ & $4370 ^{+760} _{-640}$ &  -  & $14200 ^{+380} _{-350}$ & $17270 ^{+2010} _{-2130}$ &\citet{Tan:24}\\
MilkyWay & PNG~007.6+06.9 & $990 ^{+190} _{-160}$ & $960 ^{+450} _{-360}$ &  -  & $10500 ^{+130} _{-140}$ & $12120 ^{+640} _{-590}$ &\citet{Tan:24}\\
MilkyWay & PNG~351.6-06.2 & $1260 \pm 170$ & $1140 ^{+290} _{-440}$ &  -  & $10630 ^{+160} _{-180}$ & $12760 ^{+640} _{-530}$ &\citet{Tan:24}\\
MilkyWay & PNG~352.0-04.6 & $4650 \pm 1450$ & $4090 ^{+710} _{-690}$ &  -  & $9450 ^{+100} _{-140}$ & $12230 ^{+970} _{-940}$ &\citet{Tan:24}\\
MilkyWay & PNG~352.1+05.1 & $3830 \pm 1280$ & $4610 ^{+300} _{-340}$ &  -  & $9160 ^{+40} _{-50}$ & $12790 ^{+920} _{-1390}$ &\citet{Tan:24}\\
MilkyWay & PNG~356.5-03.6 & $3440 \pm 190$ & $3350 ^{+690} _{-620}$ &  -  & $9150 ^{+210} _{-170}$ & $7000 ^{+850} _{-720}$ &\citet{Tan:24}\\
MilkyWay & PNG~358.2+03.5 & $6750 \pm 2870$ & $8990 ^{+1020} _{-820}$ &  -  & $10940 ^{+250} _{-230}$ & $13050 ^{+1380} _{-1400}$ &\citet{Tan:24}\\
MilkyWay & PNG~359.7-01.8 & $850 \pm 520$ & $1390 ^{+700} _{-520}$ &  -  & $12730 ^{+270} _{-240}$ & $14040 ^{+2140} _{-1270}$ &\citet{Tan:24}\\
MilkyWay & PNG~359.8-07.2 & $3170 \pm 880$ & $3590 ^{+590} _{-680}$ &  -  & $11660 ^{+190} _{-170}$ & $10840 ^{+1030} _{-840}$ &\citet{Tan:24}\\
\end{longtable}
$^{\rm a}$ PN names are as presented in the original references

\end{appendix}

\end{document}